\documentclass[12pt]{article}

\usepackage{amsmath}         
\usepackage{amssymb}         
\usepackage{amsfonts}        
\usepackage{pifont}          
\usepackage{graphicx}        
\usepackage{bbm}             

\usepackage{color}         
\usepackage{slashed}       
\usepackage{framed}        
\usepackage{subfig}        
\usepackage{mathrsfs}      
\usepackage{ragged2e}      
\usepackage{lipsum}        
\usepackage{paralist}      
\usepackage{multirow}      
\usepackage{cite}          

\usepackage[margin=2cm]{geometry} 

\usepackage{scalefnt}			    
\newcommand\smaller[2][0.85]{{\scalefont{#1}#2}} 

\graphicspath{{figures/}}	    

\renewcommand{\tilde}{\widetilde} 
 
\newcommand{\ket}[1]{|#1\rangle}

\numberwithin{equation}{section}

\newenvironment{institutions}[1][2em]
  {\begin{list}{}{\setlength\leftmargin{#1}\setlength\rightmargin{#1}}\item[]\RaggedRight}
  {\end{list}}
  
\renewcommand{\vec}[1]{\mathbf{#1}}
\newcommand{\Xmark}{\text{\sffamily X}}

\usepackage{tikz}
\usetikzlibrary{positioning}				    
\usetikzlibrary{calc,through}				    
\usetikzlibrary{decorations.pathreplacing}  
\usepackage{pgffor}							    
\usetikzlibrary{decorations.pathmorphing}   
\usetikzlibrary{decorations.markings}	    
\tikzset{
    vector/.style={decorate, decoration={snake}, draw},
    fermion/.style={draw=black, postaction={decorate},
        decoration={markings,mark=at position .55 with {\arrow[draw]{>}}}},
    fermionbar/.style={draw, postaction={decorate},
        decoration={markings,mark=at position .55 with {\arrow[draw]{<}}}},
    fermionnoarrow/.style={draw},
    gluon/.style={decorate, draw=
        decoration={coil,amplitude=4pt, segment length=5pt}},
    scalar/.style={dashed,draw, postaction={decorate},
        decoration={markings,mark=at position .55 with {\arrow[draw]{>}}}},
    scalarbar/.style={dashed,draw, postaction={decorate},
        decoration={markings,mark=at position .55 with {\arrow[draw]{<}}}},
    scalarnoarrow/.style={dashed,draw},
	provector/.style={decorate, decoration={snake,amplitude=2.5pt}, draw},
	antivector/.style={decorate, decoration={snake,amplitude=-2.5pt}, draw},
}

\usepackage[
	colorlinks=true,
	citecolor=black,
	linkcolor=black,
	urlcolor=blue,
	hypertexnames=false]{hyperref}

\begin{document}

\thispagestyle{empty}
\begin{center}

	{	
		\huge \bf 
		The effective theory of\\ self-interacting dark matter
	}

	\vskip .7cm
	
	\renewcommand*{\thefootnote}{\fnsymbol{footnote}}

	{	
		\bf
		Brando Bellazzini$^{a}$\footnote{\tt
		 \href{mailto:brando.bellazzini@pd.infn.it}
		 {brando.bellazzini@pd.infn.it},
		 \;
		 $^\dag$\href{mailto:mc863@cornell.edu}{mc863@cornell.edu},
		 \;
		 $^\ddag$\href{mailto:pt267@cornell.edu}{pt267@cornell.edu}
		 },
		Mathieu Cliche$^{b\dag}$,
		%
		and Philip Tanedo$^{b\ddag}$
	}  

	\renewcommand{\thefootnote}{\arabic{footnote}}
	\setcounter{footnote}{0}

	\vspace{.2cm}

\begin{institutions}[2.5cm]
\footnotesize

    
	$^{a}$ {\it  Dipartimento di Fisica e Astronomia, Universit\`a di Padova \& \textsc{infn},\\\vspace{-.2em} \hspace{0.5em} Sezione di Padova, Via Marzolo 8, I-35131 Padova, Italy} \\

	\vspace*{0.05cm}

	\hspace{0.5em} 
	{\it \textsc{sissa}, Via Bonomea 265, I-34136 Trieste, Italy} \\

	\vspace*{0.05cm}

	$^{b}$ {\it Department of Physics, \textsc{lepp}, Cornell University, Ithaca, \textsc{ny} 14853, \textsc{usa}} 

\end{institutions}

\end{center}


\begin{abstract}
\noindent 

We present an effective non-relativistic theory of self-interacting dark matter.
We classify the long range interactions and discuss how they can be generated from quantum field theories. 
Generic dark sectors can generate singular potentials. We show how to consistently renormalize such potentials and apply this to the calculation of the Sommerfeld enhancement of dark matter interactions. We explore further applications of this enhancement to astrophysical probes of dark matter including the core vs.~cusp problem.
%

%
%

\end{abstract}




\section{Introduction}

Less than a quarter of the matter density of the universe is composed of ordinary baryons.
The remaining component is called dark matter (\textsc{dm}) and has only been probed through its gravitational interactions at cosmological and astrophysical scales. 
One appealing class of \textsc{dm} candidates are weakly-interacting massive particles (\textsc{wimp}s). These are
\begin{itemize} \itemsep0em
\item stable or long-lived compared to the age of universe
\item  non-relativistic upon freeze out from thermal equilibrium in the early universe
\item  electrically neutral and weakly interacting, i.e.\ with annihilation cross sections in the pb range, so that $\Omega_\text{\textsc{DM}} h^2\approx 0.1 \mbox{ pb}/\langle \sigma v\rangle$.
\end{itemize}
These features hint at a possible link between the cosmological properties of \textsc{dm} and the mechanism for electroweak symmetry breaking.



In principle, \textsc{wimp} annihilations should still occur today in dense regions of our galaxy. The potential for this type of indirect detection has gained attention recently due to possible anomalies in the positron fluxes measured by \textsc{pamela} \cite{Adriani:2008zr},  \textsc{Fermi} \cite{FermiLAT:2011ab}  and \textsc{ams-02} \cite{Aguilar:2013qda}, and the gamma ray spectrum measured by \textsc{Fermi}  \cite{Weniger:2012tx, Su:2012ft, Rajaraman:2012db, Bloom:2013mwa, Ackermann:2012qk}.
%
%
Such signals, however, require the total \textsc{wimp} annihilation cross section to be well in excess of the thermal value.
Nevertheless, there are mechanisms to boost the annihilation rate to the level of experimental sensitivity without spoiling the relic abundance.
One possibility is that \textsc{dm} has long range self-interactions mediated by a light force carrier.
If this exchange of particles produces an attractive self-interaction, it can effectively increase the annihilation cross section because of Sommerfeld enhancement or resonance scattering 
\cite{Iengo:2009ni, 
Iengo:2009xf, 
Cassel:2009wt, 
Hisano:2004ds,
ArkaniHamed:2008qn, 
Lattanzi:2008qa,
Braaten:2013tza}.
The annihilation cross section is thus enhanced by a boost factor, $S\sigma_0$, with  $S \geq 1$, where $\sigma_0$ is the short-range annihilation cross section. 

%

More recently, self-interacting \cite{Spergel:1999mh, Dave:2000ar} 
\textsc{dm} 
has also recently been proposed as a viable solution to possible discrepancies between observations of small scale structures and the predictions from $N$-body simulations based on collisionless cold \textsc{dm}.
In particular, dwarf galaxies show flat \textsc{dm} density profiles in halo cores \cite{Rocha:2012jg,Peter:2012jh}, whereas collisionless cold \textsc{dm} predicts cusp-like profiles. In addition to this ``core vs.~cusp problem'', there is the ``missing satellites problem'' and the ``too big to fail problem,'' see e.g.~\cite{Tulin:2013teo} and references therein. 
While it is possible that these problems could be addressed by including baryonic physics to collisionless \textsc{dm} simulations \cite{Navarro:1996bv}, self-interacting \textsc{dm} offers a viable and motivated alternative scenario that is rich of interesting observational consequences \cite{Rocha:2012jg,Peter:2012jh,Chu:2011be}.

The standard approach to self-interactions and Sommerfeld enhancement is to assume an ultra-light elementary scalar or vector $\phi$ in the dark sector which mediates a force between the \textsc{dm} particles \cite{Buckley:2009in, Feng:2009hw, Tulin:2013teo}.
In this paper we take a more agnostic approach; we construct an effective theory that only assumes rotationally invariant self-interactions in the dark sector. One can classify the possible potentials in terms of the \textsc{dm} mass $m_{\chi}$, spin $\vec{s}$, transferred momentum  $\vec{q}$, and relative velocity $\vec{v}$.
We work at the leading order in the exchanged momentum and velocity which is an excellent approximation for cold \textsc{dm}.
For example, we show in Section~\ref{sec:eff:pot} that the most general long-range $P$- and $T$- symmetric potential between two \textsc{dm} particles of arbitrary spin, is
\begin{align}
\label{eq:potentialCP}
V^{P,T}_{\mathrm{eff}}= \frac{1}{4\pi r} \left[ \tilde{g}_1(r)+\tilde{g}_2(r) (\vec{s}_1\cdot \vec{s}_2) + \frac{\tilde{g}_3(r)}{\Lambda^2 r^2}\left(3\vec{s}_1\cdot \hat{r} \,\,\,\vec{s_2}\cdot\hat{r}-\vec{s}_1 \cdot \vec{s}_2\right) +\frac{\tilde{g}_{7,8}(r)}{\Lambda r}(\vec{s}_1\pm \vec{s}_2)(\hat{r}\times \vec{v})\right] 
\end{align}
where $\tilde{g}_i(r)$ are arbitrary functions that depend only on the the \textsc{dm} separation, 
and $\Lambda$ is the characteristic interaction scale of the dark sector that we take much larger than the mediator mass. 
At scales where the mediator mass can be neglected and the theory is weakly coupled, the couplings $\tilde{g}_i$ freeze to constants, $\tilde{g}_i(r)\rightarrow g_i $.
%

Strongly interacting mediators in the dark sector can generate singular potentials  through non-standard propagators, see e.g.~\cite{Georgi:2007ek,Georgi:2009xq}. 
Notice, however, that even weakly coupled models can generate potentials that are more singular than the $1/r^2$ centrifugal barrier at short distances.
For example, dark matter interactions mediated by a light pseudo-scalar produce a $g_3$ term in the potential (\ref{eq:potentialCP}) which goes like $1/r^3$. This can be generated, for example, by Goldstone bosons \cite{Bellazzini:2011et}.
%
Another example is \textsc{dm} with dipole interactions generated by charged states. These produce a $g_3$ term in the potential. Models based on these magnetic dipole interactions were recently proposed \cite{Chang:2010en} as a way to resolve discrepancies between tentative signals in direct detection experiments.
More exotic potentials can be generated by the loop-level exchange of composite operators made of light fields \cite{Ferrer:1998rw,Hsu:1992tg,Feinberg:1989ps, Dobrescu:2006au}.
%
%
Table~\ref{table:operators} shows examples of weakly coupled models, preserving $P$ and $T$, that generate the various $g_i$ in (\ref{eq:potentialCP}). 
%
%
\begin{table}
\begin{equation*}
\begin{array}{cccccc}
\hline
\text{Interaction} & g_1 & g_2 & g_3 & g_7 & g_8\\
\hline
\bar{\chi}\chi \varphi & \checkmark & \Xmark & \Xmark  & \checkmark &\Xmark \\
\bar{\chi}\gamma^5\chi \varphi &  \Xmark & \Xmark & \checkmark & \Xmark & \Xmark \\
i\bar{\chi}\gamma^\mu\gamma^5\chi \partial_\mu\varphi & \Xmark & \Xmark & \checkmark & \Xmark & \Xmark\\
\bar{\chi}\gamma^\mu\chi A_\mu & \checkmark & \Xmark & \Xmark & \checkmark &\Xmark \\
i \bar{\chi}\gamma^5\gamma^\mu\chi  A_\mu & \Xmark & \Xmark & \checkmark & \Xmark & \Xmark \\
i\bar{\chi}\sigma^{\mu\nu}\chi F_{\mu\nu}  & \Xmark &  \Xmark & \checkmark & \Xmark & \Xmark \\
\hline
\end{array}
\end{equation*} 
\caption{
Leading order $P$- and $T$-preserving long-range static potentials in (\ref{eq:potentialCP}) from massless real scalar $\varphi$, vector gauge boson $A_\mu$, or field strength $F_{\mu\nu}=\partial_{\left[\mu\right.} A_{\left.\nu\right]}$ mediators. Observe that $g_2$ is not generated in the massless limit. $g_{8}$ is not generated because of the spin conservation in $CP$-symmetric theories of spin-$\frac{1}{2}$ \textsc{dm}. See Table~\ref{table:operators2} and \ref{table:operators3}  for more details. }
\label{table:operators}
\end{table}

Such singular potentials must be regularized at short distances and then renormalized by requiring that low-energy observables are cutoff independent.  
We carry out this renormalization program making possible to extract physical predictions from singular potentials generated by \textsc{dm} self-interactions. In particular, we calculate the  Sommerfeld enhancement from a $1/r^3$ potential, extending the analysis in~\cite{Bedaque:2009ri} by including wavefunction renormalization
\footnote{
%
We note that wavefunction renormalization is essential for Sommerfeld enhancement to be cutoff independent. The numerical results in Section~\ref{sec:numerical} match \cite{Bedaque:2009ri} within an order of magnitude for a specific choice of renormalization conditions.
}. We plot the elastic scattering cross section as a function of the velocity and the mass near the resonance region where the boost factor is large.
Astrophysical systems at various  scales, from clusters to dwarf galaxies with velocity ranging from $v\sim 10^{-3}$ and $v\sim 10^{-5}$, provide constraints on the \textsc{dm} self-interactions and hence the Sommerfeld enhancement \cite{Buckley:2009in, Feng:2009hw, Tulin:2013teo}. 
While we leave an investigation of how these bounds may be adapted to singular potentials for future work, we point out that the formalism presented here may be useful to avoid these constraints because of the velocity dependence of the elastic cross section. 

Even though Sommerfeld enhancement is typically relevant only for $s$-wave annihilations due to the centrifugal barrier, the self-interacting \textsc{dm} potential (\ref{eq:potentialCP}) does not generically conserve orbital angular momentum $\vec{L}^{2}$. Interaction channels with different orbital angular momenta, $\ell$, can be coupled. This explains why the $g_3$ term in (\ref{eq:potentialCP}), which would be averaged to zero because of isotropy of $\ell=0$ states, can still be relevant for Sommerfeld enhancement in $\Delta\ell=2$ transitions \cite{Bedaque:2009ri}. 
Moreover, spin-spin  interactions with $g_3\neq0$ in (\ref{eq:potentialCP}) may generate macroscopic long range interactions when the \textsc{dm} spins are polarized (in average) \cite{Dobrescu:2006au}, a condition that on galaxy scales may be plausible for these $\vec{L}$-violating interactions.

This paper is organized as follows. 
In Section~2 we derive an effective long-range, non-relativistic potential for self-interacting dark matter at leading order in \textsc{wimp} velocity. In Section~3 we present a procedure to renormalize singular potentials and apply this to the calculation of the physical, cutoff-independent Sommerfeld enhancement. In Sections~4 and 5 we present numerical results for a $1/r^3$ potential and discuss the types of astrophysical bounds that such an analysis may be applied to. We conclude in Section~6 and include appendices reviewing the standard procedure for calculating Sommerfeld enhancement for non-singular potentials and a convenient square well approximation for singular potentials.

\section{Effective long-range potential}
\label{sec:effectivesect}

The elastic scattering amplitude $\mathcal{M}$ from rotationally invariant \textsc{dm} self-interactions is a scalar function of the spins $\vec{s}_i$, exchanged momentum $\vec{q}$, and relative velocity $\vec{v}$.
It is often convenient to use the Hermitian operators $i \vec{q}$ and the velocity transverse to the momentum transfer, 
\begin{align}
\vec{v}_\perp=\vec{v}-\frac{\vec{q}(\vec{q}\cdot\vec{v})}{\vec{q}^2}=\vec{v}+\vec{q}/m_\chi
\end{align}
where the last equality follows from the four-momentum conservation.  

In the center of mass frame, the elastic scattering amplitude is
\begin{equation}
\label{eq:Mamplitude}
\mathcal{M}=\frac{-1}{\vec{q}^{2}+m_\phi^2}\sum_i g_{i}(\vec{q}^{2}/\Lambda^2,\vec{v}_\perp^2)\mathcal{O}_i( \vec{s}_{j}\cdot i\vec{q}/\Lambda,  \vec{s}_{j}\cdot \vec{v}_\perp,\vec{s}_1\cdot\vec{s}_2)
\end{equation} 
where $\Lambda$ is the heavy scale of the dark sector, e.g.\ the \textsc{dm} mass $m_\chi$, and $\mathcal{O}_i$ are the spin matrix elements. 
We explicitly pull out a factor associated with the propagator for the light force carrier with mass $m_\phi^2\ll \vec{q}^{2}\ll \Lambda^2$ which acts as an infrared (\textsc{ir}) regulator at large distances. 
Further, we only consider the leading term in the exchanged momentum $\vec{q}/\Lambda$ and \textsc{dm} velocities, which we assume to be small $v\,, v_\perp\ll 1$. This is a good approximation for cold \textsc{dm} in the phenomenologically interesting regime from dwarf galaxy scales $v\sim 10^{-5}$ to freeze out $v\sim 0.3$. 
This type of non-relativistic effective theory was recently applied to the direct detection of dark matter in \cite{Fan:2010gt, Fitzpatrick:2012ix}.
In order to conserve \textsc{dm} energy (and the total angular momentum) we assume that mediator bremsstrahlung is kinematically suppressed, $m_\chi \vec{v}^2\ll m_\phi$. In other words, we work in the regime 
\begin{equation}
\label{eq:regime}
\vec{v}^4\ll \frac{m_\phi^2}{m_\chi^2} \ll \frac{\vec{q}^2}{m_\chi^2}\sim \vec{v}^2\,.
\end{equation} 
We assume mediators with spin less than 2 since the longitudinal components of massive particles with higher spins spoil the derivative expansion at scales comparable with their mass, $\vec{q}\sim m_\phi$.
%

\subsection{Rotationally invariant non-relativistic operators}

Under parity and time reversal velocities, spins and momentum, transform as
\begin{align}
P:\,\, i\vec{q}&\rightarrow  -i\vec{q}\,,\qquad \vec{s}\rightarrow    +\vec{s}\,,\qquad \vec{v}_\perp \rightarrow -\vec{v}_\perp\,, \\
T:\,\,  i\vec{q}&\rightarrow   +i\vec{q}\,,\qquad \vec{s}\rightarrow   -\vec{s}\,,\qquad \vec{v}_\perp \rightarrow -\vec{v}_\perp\,.
\end{align}
In turn, one can build the following invariant parity-even operators
\begin{align}
\mathcal{O}_{1\phantom{,0}}= & \phantom{+} 1
\label{eq:op:1}
\\
\mathcal{O}_{2\phantom{,0}}= & \phantom{+} \vec{s}_1\cdot \vec{s}_2 
\label{eq:op:2}
\\
\mathcal{O}_{3\phantom{,0}}= & - \frac{1}{\Lambda^2}(\vec{s}_1\cdot \vec{q})(\vec{s}_2\cdot \vec{q})
\label{eq:op:3}
\\
\mathcal{O}_{4\phantom{,0}}= & \phantom{+} (\vec{s}_1\cdot \vec{v}_\perp)(\vec{s}_2\cdot \vec{v}_\perp)
\label{eq:op:4}
\\
\mathcal{O}_{5,6}=&-\frac{i}{\Lambda}\left[(\vec{s}_1\cdot \vec{q})(\vec{s}_2\cdot \vec{v}_\perp)\pm (\vec{s}_1\cdot \vec{v}_\perp)(\vec{s}_2\cdot \vec{q})\right]
\label{eq:op:78}
\\
\mathcal{O}_{7,8}=& -\frac{i}{\Lambda}\left[(\vec{s}_1\pm \vec{s}_{2})\cdot (\vec{q}\times \vec{v})\right]\,,\label{eq:op:78}
\end{align}
where spin wavefunctions are suppressed for simplicity. Operators $\mathcal{O}_{5,6}$ respect parity but break time reversal. 
In the following we discard $\mathcal{O}_4$ because it is only generated by spin-2 mediators \cite{Fitzpatrick:2012ix}.
%
Relaxing parity invariance introduces eight additional operators \cite{Dobrescu:2006au}: four of those respect time reversal or, equivalently, $CP$
\begin{align}
\mathcal{O}_{9\phantom{0,00}}=&\,-\frac{1}{\Lambda} (\vec{s}_1\times \vec{s}_{2})\cdot i\vec{q} \,,
\label{eq:O9}
\\
\mathcal{O}_{10,11}=&\,\phantom{+} (\vec{s}_1\pm \vec{s}_{2})\cdot \vec{v}_\perp \,, 
\\
\mathcal{O}_{12\phantom{,00}}=&\, -\frac{i}{\Lambda}[\vec{s}_1\cdot (\vec{q}\times \vec{v})](\vec{s}_2\cdot \vec{v}_\perp) +\frac{i}{\Lambda}[\vec{s}_2\cdot (\vec{q}\times \vec{v})](\vec{s}_1\cdot \vec{v}_\perp)\,,
\label{eq:12}
\end{align}
while other four break both $P$ and $CP$
\begin{align}
\label{eq:13}
\mathcal{O}_{13,14}=&\,- \frac{1}{\Lambda} (\vec{s}_1\pm \vec{s}_{2})\cdot i\vec{q}\,, 
\\
\mathcal{O}_{15\phantom{,00}}=&\, \phantom{+} (\vec{s}_1\times \vec{s}_{2})\cdot \vec{v}_\perp \,,
\\
\mathcal{O}_{16\phantom{,00}}=&\, -\frac{1}{\Lambda^2}(\vec{s}_2\cdot \vec{q})[\vec{s}_1\cdot (\vec{q}\times \vec{v})] +\frac{1}{\Lambda^2}(\vec{s}_1\cdot \vec{q})[\vec{s}_2\cdot (\vec{q}\times \vec{v})]\,.
\label{eq:16}
\end{align}
%
%
Observe that self-conjugate \textsc{dm} is symmetric under the exchange $1\leftrightarrow 2$. This is equivalent to invariance under $(\vec{q},\vec{v},\vec{s}_1) \leftrightarrow (-\vec{q},-\vec{v},\vec{s}_2)$, which forbids $\mathcal O_{6,8,10,12,13,16}$.

\subsection{The general effective potential}
\label{sec:eff:pot}

%
%
%

A more general approach is to replace the free propagator with a general two point function in (\ref{eq:Mamplitude}). This may include arbitrary negative powers of $\vec{q}^2$ from non-local interactions mediated by light states that have been integrated out. In an integral representation, the amplitude is
%
\begin{equation}
\mathcal{M} 
=
-\int_0^\infty d\mu^2 
\frac{\rho(\mu^2)}{\vec{q}^{2}+\mu^2}
\sum_i g_{i}(\vec{q}^2/\Lambda^2,\vec{v}_\perp^2)\mathcal{O}_i(\vec{v}_j\cdot i\vec{q}/\Lambda,\vec{s}_{i}\cdot \vec{v}_\perp,\vec{s}_1\cdot\vec{s}_2)
\label{eq:Mamplitude:gen}
\end{equation} 
where $\rho(\mu^2)$ is the spectral density of the theory which provides a common language to describe weakly and strongly coupled models.
The standard propagator is recovered when $\rho(\mu^2)=\delta(\mu^2-m_{\phi}^2)$.

Since the couplings always appear with the mediator's propagator, we can make the replacement $g_i(\vec{q}^2/\Lambda^2,\vec{v}_\perp^2)=g_i(-\mu^2/\Lambda^2,\vec{v}_\perp^2)$ after neglecting short-range interactions such as $\delta$-functions. 
Moreover, for light mediators, the spectral density only has support for $\mu^2 \ll m_\chi^2,\Lambda^2$ so that we may further write $g_i(\vec{q}^2/\Lambda^2,\vec{v}_\perp^2)\simeq g_i(0,0)\equiv g_i$ unless this order vanishes.
In such a case one should go to the first non-vanishing order, $g_i\rightarrow (-\mu^2/\Lambda^2)^n g^{(n)}_i/n!$.
We have also dropped the velocity dependence because it does not provide the leading contribution unless one fine tunes the coefficients of the \textsc{uv} operators to cancel the velocity-independent contributions  \cite{Fan:2010gt,Fitzpatrick:2012ix}.
%
%

Taking the Fourier transform of the scattering amplitude with respect to $\vec{q}$ , one obtains the long-range effective potential as a function of the relative distance $\vec{r}$ and velocity $\vec{v}$.
For example, $P$- and $T$-symmetric interactions result in an effective long-range potential
\begin{align}
\label{eff:potential:PandT}
V^{P,T}_{\mathrm{eff}}= & \frac{1}{4\pi r}\left\{ \tilde{g}_1(r)
+\tilde{g}_2(r) (\vec{s}_1\cdot \vec{s}_2) 
 +\frac{\tilde{g}_3(r)}{\Lambda^2 r^2}\left[ 3\vec{s}_1\cdot \hat{r} \,\,\,\vec{s_2}\cdot\hat{r}-\vec{s}_1 \cdot \vec{s}_2\right]
+\frac{\tilde{g}_{7,8}(r)}{\Lambda r}(\vec{s}_1\pm \vec{s}_2)(\hat{r}\times \vec{v})\right\}
\end{align}
where $\tilde{g}_i(r)$  are integrals of the Yukawa factor over the spectral density
\begin{align}
\label{V0spectral}
\tilde{g}_1(r)=& \phantom{g_{7,8}} \int_0^\infty d\mu^2 \rho(\mu^2) e^{-\mu r} \left(g_1-g^{(1)}_1\frac{\mu^2}{\Lambda^2}\right)\\
\tilde{g}_2(r)= & \phantom{g_{7,8}} \int_0^\infty d\mu^2 \rho(\mu^2) e^{-\mu r} \left[g_2+\left(\frac{g_3}{3}-g_2^{(1)}\right)\frac{\mu^2}{\Lambda^2}\right]\\
\tilde{g}_3(r)=& \, \phantom{_{,0}} g_{3} \int_0^\infty d\mu^2 \rho(\mu^2) e^{-\mu r} \left[1+\mu r+\frac{1}{3}(\mu r)^2\right]\\
\tilde{g}_{7,8}(r)= & \, g_{7,8} \int_0^\infty d\mu^2 \rho(\mu^2) e^{-\mu r}  \left(1+\mu r \right)
\label{dofr}
\end{align}
It is understood that working at the leading non-vanishing order,  $g_{3,7,8}$ and $g^{(1)}_{1,2}$ should always be dropped unless the  $\mathcal{O}(\vec{q}^0)$ terms like $g_{1,2}$ are vanishing or suppressed.  
For weakly coupled dark sectors, $\tilde{g}_{1,2}$ are the usual exponential factors while $\tilde{g}_{3,7,8}$ carry additional polynomial corrections in the mediator mass.
In general these functions have an arbitrary $r$ dependence, as expected when the mediator is a composite operator.
A simple example is a four-fermi operator between spin-$\frac{1}{2}$ \textsc{dm} particles $\chi$ and a massless neutrino-like species $\nu$, that is $\mathcal{L}=\sqrt{\alpha}[\bar{\nu}\gamma_{\mu}(1-\gamma_5)\nu][\bar{\chi}\gamma_{\mu}(a-b\gamma_5)\chi]$.  The mediator is a composite operator made of two light fermions. It generates a singular potential at the loop level \cite{Hsu:1992tg}, $\tilde{g}_{i=1,2}\propto 1/r^{4}$ and $\tilde{g}_3\propto \Lambda^2/r^2$.
Note that the spin structure of the potential is fixed by the quantum numbers of the light mediator.  

Note that for spin-$\frac{1}{2}$ \textsc{dm} the particle--antiparticle potential must have $g_8=0$  since $CP$
 corresponds to a factor $(-)^{S+1}$ and thus implies the conservation of total spin $\vec{S}^2=\left(\vec{s}_1+\vec{s}_2\right)^2$ which can only take values $0$ and $1$.
In this case, it is convenient to express the potential in the following form
\begin{align}
 V^{(s_i=1/2)}=& \frac{1}{4\pi r}
 \left\{
 \left(\tilde{g}_1(r)-\frac{3}{4}\tilde{g}_2(r)\right) + 
 \frac{1}{2}\tilde{g}_2(r)\vec{S}^{\,2}
 +
 \frac{\tilde{g}_3(r)}{2\Lambda^2 r^2} 
 \left[
     3(\vec{S}\cdot \hat{r})^2 -\vec{S}^{\,2}\right]  +
\frac{2\tilde{g}_{7}(r)}{m_\chi \Lambda\, r^2}\vec{S}\cdot\vec{L}
\right\},
\label{eq:potential:parity}
\end{align}
where $\vec{L}=\vec{r}\times \vec{p}$ is the orbital angular momentum and $\vec{p}=m_\chi\vec{v}/2$ is the conjugate momentum, $[\vec{r}^i,\vec{p}^j]=i\delta^{ij}$.  
%

At large distances,  but smaller than the mediator Compton wavelength, $\Lambda^{-1}\ll r \ll \mu^{-1}$,  the functions $\tilde{g}_{i}(r)$ become constants and the potential simplifies even further:
\begin{align}
\label{eq:V:eff:parity}
V^{P,T}_{\mathrm{eff}}= \frac{1}{4\pi r} \left[ g_1+g_2 (\vec{s}_1\cdot \vec{s}_2) + \frac{g_3}{\Lambda^2 r^2}\left(3\vec{s}_1\cdot \hat{r} \,\,\,\vec{s_2}\cdot\hat{r}-\vec{s}_1 \cdot \vec{s}_2\right) +\frac{g_{7,8}}{\Lambda r}(\vec{s}_1\pm \vec{s}_2)(\hat{r}\times \vec{v})\right]  \,.
 \end{align}%
This is the regime where Sommerfeld enhancement may be effective because the interaction is still long-range compared to the short distance annihilation processes that take place at $r\sim\Lambda^{-1}$.

The expressions for the potentials that break $P$ but respect $T$ are presented in Appendix~\ref{app:list}.

\subsection{Weakly coupled examples}
\label{sec:weakly:coupled:examples}

\begin{table}
\begin{center}
\begin{tabular}{llccc}
\hline
mediator 
&  interaction
& $\dfrac{1}{r}$ 
& $\dfrac{1}{r}\left(\vec{s}_1 \cdot \vec{s}_2\right)$ 
& $\dfrac{1}{r^3}\left[3 \left(\vec{s}_1\cdot \hat{r}\right)  \left(\vec{s_2}\cdot\hat{r}\right)-\vec{s}_1 \cdot \vec{s}_2\right]$ \\
 \noalign{\smallskip}
\hline
\noalign{\smallskip}
scalar 
& $ \lambda_s \bar\chi\chi \varphi $
&  $-\lambda_s^2$ 
& 0 
& 0
\\
\noalign{\smallskip}
pseudoscalar &
$ i\lambda_p \bar\chi\gamma^5\chi \varphi $ 
& 0 
& $\displaystyle \frac{\lambda_p^2 m_\varphi^2}{3 m_\chi^2}$ 
& $\displaystyle \frac{\lambda_p^2}{m_\chi^2} h(m_\varphi,r)$
\\
\noalign{\smallskip}
Goldstone &
$\displaystyle \frac{1}{f} \bar\chi\gamma^\mu \gamma^5 \chi \partial_\mu \varphi $ 
& 0 
& $\displaystyle \frac{4m_\varphi^2}{3f^2}$ 
& $\displaystyle \frac{4}{f^2} h(m_\varphi,r)$ 
\\
\noalign{\smallskip}
vector &
$ \lambda_v \bar\chi\gamma^\mu\chi A_\mu $ 
& $\displaystyle \pm \lambda_v^2\left(1+\frac{m_A^2}{4m_\chi^2}\right)$ 
& $\displaystyle \pm\frac{2\lambda_v^2 m_A^2}{3m_\chi^2}$
& $\displaystyle \mp\frac{\lambda_v^2}{m_\chi^2} h(m_A,r)$ 
\\
\noalign{\smallskip}
axial vector &
$ \lambda_a \bar\chi\gamma^5\gamma^\mu\chi A_\mu $ 
& 0 
& $\displaystyle-\frac{8\lambda_a^2}{3}\left(1-\frac{m_A^2}{8m_\chi^2}\right)$
& $\displaystyle \lambda_a^2 \left(\frac{1}{m_\chi^2}+\frac{4}{m_A^2}\right) h(m_A,r)$ 
\\
\noalign{\smallskip}
field strength &
$\displaystyle \frac{i}{2\Lambda} \bar\chi \sigma^{\mu\nu}\chi F_{\mu\nu} $ 
& 0 
& $\displaystyle \mp \frac{2m_A^2}{3\Lambda^2}$
& $\displaystyle  \pm\frac{1}{\Lambda^2} h(m_A,r)$ \\
\noalign{\smallskip}
\hline
\end{tabular}
\end{center}
\caption{Parity-preserving particle--(anti-)particle (upper/lower sign) long-range, static potentials  from scalar $\varphi$, gauge boson $A_\mu$, and field strength $F_{\mu\nu}=\partial_{\left[\mu\right.} A_{\left.\nu\right]}$ mediators. 
Here $\sigma^{\mu\nu}=\frac{i}{4}[\gamma^\mu,\gamma^\nu]$ and $h$ is defined in (\ref{eq:silly:h:eq}). 
Each term implicitly carries a Yukawa factor ${e^{-m_\phi r}}/{4\pi}$.
Observe that the long-range $\vec{s}_1\cdot\vec{s}_2$ is always suppressed by the mediator mass since $\lambda_a = m_A/f$.
}
\label{table:operators2}
\end{table}

As an example, consider a dark sector with a weakly coupled, light scalar or vector mediator $\phi$ with interactions $\lambda \mathcal O^\text{QFT}$ in Table~\ref{table:operators2}. These generate a static potential
\begin{align}
\sum_i\lambda_i \mathcal O_i^\text{QFT} &\longrightarrow
V^P_{\mathrm{s}}=
\left[
g_1
+ 
g_2 (\mathbf{s}_1\cdot \mathbf{s}_2)
+
\frac{g_3}{\Lambda^2 r^2} h(m_\phi,r)
\left[3 \left(\vec{s}_1\cdot \hat{r}\right)  \left(\vec{s_2}\cdot\hat{r}\right)-\vec{s}_1 \cdot \vec{s}_2\right] 
\right]
\frac{e^{-m_\phi r}}{4\pi r},
\label{eq:V:with:pot}
\end{align}
%
where $h$ encodes the dependence on the mediator mass, 
\begin{align}
h(m_\phi,r)=\left(1 + m_\phi r + \frac{m_\phi^2 r^2}{3} \right).
\label{eq:silly:h:eq}
\end{align}
Table~\ref{table:operators2} gives the contributions to each of the coefficients on the right-hand side of (\ref{eq:V:with:pot}) coming from the corresponding types of \textsc{qft} interactions.

Note that the Dirac \textsc{dm} mass $m_\chi$ breaks axial symmetry so that the limit of a massless axial gauge boson mediator is  consistent at finite $m_\chi$ only when chiral symmetry is broken spontaneously at a scale $f$ so that $m_A= \lambda_a f$.  In this case the transverse components decouple, $\lambda_a = m_A/f\rightarrow 0$, and only the longitudinal modes contribute to the amplitude with coupling $1/f$, matching the result from Goldstone boson exchange.

Table~\ref{table:operators3} gives the long-range, non-static potential contributions from massless scalars and gauge bosons.
 The $\vec{v}_\perp^2$ contribution generates a $\sim 1/r(\vec{v}^2+\hat{r}(\hat{r}\cdot\vec{v})\vec{v})$ in position space which can be neglected because it is always subleading.
Pseudo-scalar, axial-vector and field strength mediators, give vanishing non-static, long-range potentials at this order.
Note that these potentials generically need  to be complemented by the relativistic corrections to the kinetic energies, $\vec{p}^2/m_\chi^2\left(1-\vec{p}^2/(4m_\chi^2)+\ldots\right)$. 
In the following sections we neglect these corrections to the kinetic energy since we checked that their contribution is very small. 

%
\begin{table}
\begin{center}
\begin{tabular}{llcc}
\hline
mediator 
&  interaction
& $\dfrac{1}{4\pi r}\left[\vec{v}^2+\hat{r}(\hat{r}\cdot\vec{v})\vec{v}\right]$ 
& $\dfrac{1}{4\pi r^2}\left(\hat{r} \times \vec{v}\right)\cdot (\vec{s}_1+\vec{s}_2)$  \\
 \noalign{\smallskip}
\hline
\noalign{\smallskip}
scalar 
& $ \lambda_s \bar\chi\chi \varphi $
&  $\displaystyle -\frac{\lambda_s^2}{8}$ 
& $\displaystyle \frac{\lambda_s^2}{4m_\chi}$ 
\\

\noalign{\smallskip}
vector &
$ \lambda_v \bar\chi\gamma^\mu\chi A_\mu $ 
& $\displaystyle \pm \frac{\lambda_v^2}{8}$ 
& $\displaystyle \mp\frac{3\lambda_s^2}{4m_\chi}$
\\
\noalign{\smallskip}
\hline
\end{tabular}
\end{center}
\caption{Parity-preserving particle--(anti-)particle (upper/lower sign) long-range, non-static potentials from massless scalars $\varphi$ and gauge bosons $A_\mu$. Long-range contributions from pseudo-scalars, axial vectors and field strength vanish for massless mediators. 
%
%
}
\label{table:operators3}
\end{table}

\section{Renormalization of singular potentials and Sommerfeld enhancement}

The potential $V^{P, T}_{\mathrm{eff}}$ in (\ref{eff:potential:PandT}) represents the most general long-range interactions between \textsc{dm} particles that preserve parity and time reversal. 
A standard method for calculating the Sommerfeld enhancement for the non-singular Coulomb and Yukawa potentials is presented in \cite{Iengo:2009ni} and reviewed in Appendix~\ref{app:origin:of:sommerfeld}.
In practice, one determines the boost factor by solving a Schr\"odinger-like equation with the proper boundary conditions. 
%
%
%
However, since the terms in $V^{P,T}_{\mathrm{eff}}$ are typically very singular, the usual calculations for the boost factor will generically fail.  In this section we show how to overcome these problems by renormalizing the Schr\"odinger equation. Since a full numerical solution can be computationally intensive for singular potentials, we also provide an algebraic algorithm to estimate the Sommerfeld enhancement for general potentials in Appendix~\ref{app:box}. 



\subsection{Wilsonian treatment of divergences}

Potentials that go to infinity faster than $1/r^2$ at the origin are called singular \cite{Frank:1971xx} and generically arise in dark sectors with spinning \textsc{dm} and/or with some strong dynamics.  
%
%
%
The occurrence of unphysical behavior originating from the infinitely large energies of such potentials are analogous to the infinities of quantum field theory (\textsc{qft}). These inconsistencies arise when one extrapolates a long-range potential to arbitrarily short distances where ultraviolet physics should be taken into account.
In fact, the Schr\"odinger equation can be renormalized by adopting the Wilsonian renormalization group (\textsc{rg}) methods of \textsc{qft} \cite{Lepage:1997cs}: the singular potential is regulated at a short distance $a$ and augmented with a series of local operators that parametrize the unknown \textsc{uv} physics,
\begin{equation}
\label{Vwilson}
V(r)\longrightarrow 
V(r)\theta(r-a)+c_0(a)\delta^3(r)+c_2(a)a^2\nabla^2\delta^3(r)+\ldots
\end{equation}
The short-distance part of this effective potential is a derivative expansion that can be truncated to the desired order as long as the typical momenta $q$ are much smaller than the cutoff scale $\Lambda=a^{-1}$. 
This given order in $q$ determines the finite set of coupling constants $c_{i}(a)$ which can be determined by low-energy data.

\subsection{Renormalized potential}

Singular potentials diverge at the origin so that further care is required to impose boundary conditions.
The Schr\"odinger equation for an $\ell$-wave state is conveniently expressed using the dimensionless coordinate $x= p r$, the product of the dark matter relative momentum and separation:
\begin{align}
	-\Phi_{p,\ell}''(x)
	+ \left( \mathcal{V}(x)+\frac{\ell(\ell+1)}{x^2}-1\right)
	\Phi_{p,\ell}(x)
	&=0,
	\label{eq:Schrodinger:dimless}
\end{align}
where the dimensionless potential is 
rescaled by the momentum $p$ and reduced mass $M=m_{\chi}/2$,
\begin{align}
\mathcal V(x) = \frac{2M}{p^2} V\left(\frac{x}{p}\right).
\end{align}
We regulate the potential at $x_\text{cut}=ap$ with a square well of height $\mathcal V_0$ encoding the \textsc{uv} data of the relativistic completion,
\begin{align}
	\mathcal{V}_\text{reg}(x)
	=
	\mathcal{V}(x) \, \theta(x-x_\text{cut})
	+ 
	\frac{1}{x_\text{cut}^2}\mathcal{V}_0\,\theta(x_\text{cut}-x).
	\label{eq:Vreg}
\end{align}
In practice, we simulate the local counter-terms with a short-distance square well potential which makes the calculations much easier  \cite{Beane:2000wh}. We stress, however, that any other choice or deformation  of the counter-terms is allowed and physically equivalent as long as it changes only the UV behavior of the interactions \cite{Lepage:1997cs}.

Observe that the centrifugal barrier is left uncut since it is non-singular and unrelated to the \textsc{uv} physics.   
Once $\mathcal V_0$ is known, one may integrate the Schr\"odinger equation subject to the usual boundary condition at zero
\begin{align}
\lim_{x\rightarrow 0}	\Phi_{p,\ell}(x) = x^{\ell+1},
\label{eq:schrodinger:BC:at:0}
\end{align}
and then extract the Sommerfeld enhancement from the asymptotic solution. In the regulated region $x<x_\text{cut}$, the Schr\"odinger equation can be solved explicitly in the approximation $x_\text{cut}\ll 1$,
%
\begin{eqnarray}
\Phi_{p,\ell}(x<x_\text{cut})
=
\Gamma\left(\ell+\frac{3}{2}\right)
\left(
\frac{
    2x_\text{cut}
}{
    \mathcal{V}_0^{1/2}
}
\right)^{\ell+1/2}
x^{1/2}\,
J_{\ell+1/2}\left(\mathcal{V}_0^{1/2} \frac{x}{x_\text{cut}}\right).
\label{eq:schro:sol:x:less:xcut}
\end{eqnarray}

 The value $\mathcal V_0$ that appears in the Schr\"odinger equation is determined by requiring that a low energy observable is independent of the particular choice of the cutoff, $x_\text{cut}$. It is thus meaningful to define $\mathcal V_0(x_\text{cut})$ with respect to the value of a physical observable, which can be conveniently chosen to be the scattering phase $\delta_\ell$ of the elastic dark matter scattering process that generates this enhancement.

For the region $x>x_\text{cut}$, recall that the general solution to the Schr\"odinger equation is a linear combination of two independent solutions,
\begin{eqnarray}
\Phi_{p,\ell}(x>x_\text{cut})=Af(x)+Bg(x).
\label{eq:schro:sol:gen:x:gtr:xcut}
\end{eqnarray}
Asymptotically far from the origin, these independent solutions are combinations of sines and cosines. The scattering phase is related to the shift in the argument when the asymptotic solution is written as a pure sine. Thus the $\delta_\ell$ has a one-to-one relation to the ratio $A/B$. In this way $A/B$ contains the \textsc{uv} data that can be measured in a low energy observable, the scattering phase shift.

In order to determine $\mathcal V_0(x_\text{cut})$ subject to a fixed scattering phase, we may match the logarithmic derivatives of the two piecewise solutions at $x_\text{cut}$. Comparing (\ref{eq:schro:sol:x:less:xcut}) with (\ref{eq:schro:sol:gen:x:gtr:xcut}), 
\begin{eqnarray}
    -\frac{\ell}{x_\text{cut}}
    +
    \frac{
        \mathcal{V}_{0}^{1/2}(x_\text{cut})
    }{
        x_\text{cut}
    }
    \frac{
        J_{\ell-1/2}\left(\mathcal{V}_{0}^{1/2}(x_\text{cut})\right)
        }{J_{\ell+1/2}\left(\mathcal{V}_{0}^{1/2}(x_\text{cut})\right)}
      &=&\frac{\frac{A}{B}f'(x_\text{cut})+g'(x_\text{cut})}{\frac{A}{B}f(x_\text{cut})+g(x_\text{cut})}.
\label{eq:matching:at:xcut}
\end{eqnarray}
Observe that matching the logarithmic derivative gives an expression that depends on $A/B$ which is cutoff independent and directly related to our low-energy observable \cite{Beane:2000wh}. 
Once $\mathcal V_0(x_\text{cut})$ is determined, (\ref{eq:Vreg}) is the correct non-singular low-energy potential for the problem with the given cutoff. 
%
%

Due to the oscillatory nature of the Bessel function, there can be multiple solutions to the transcendental equation (\ref{eq:matching:at:xcut}). These solutions are physically equivalent. To simplify our calculations we choose the first quadrant so that $\mathcal{V}_0(x_\text{cut})$ can take values in the range $(-\infty,\mathcal{V}_\text{max})$ where $\mathcal{V}_\text{max}$ is given by the first positive solution of 
\begin{eqnarray}
J_{\ell+1/2}\left(\mathcal{V}_\text{max}^{1/2}\right)&=&0
\end{eqnarray}
For $\ell=0$, $\mathcal{V}_\text{max}=\pi^2$.


\subsection{Wavefunction renormalization}

Since  $\mathcal V_\text{reg}$ in (\ref{eq:Vreg}) is manifestly non-singular, one may proceed to solve the Schr\"odinger equation (\ref{eq:Schrodinger:dimless}) subject to (\ref{eq:schrodinger:BC:at:0}) following the procedure outlined in Appendix~\ref{app:origin:of:sommerfeld}. The resulting Sommerfeld enhancement, $S^{(0)}$, appears to depend on the choice of $x_\text{cut}$. This residual cutoff dependence is not physical and is removed by including wavefunction renormalization, $Z_{\ell}$:
\begin{eqnarray}
S_{\ell}=Z_\ell S^{(0)}_\ell
\end{eqnarray}
$Z_\ell$ is fixed by using the observation that at relativistic speeds the Sommerfeld enhancement factor should go to one, 
\begin{eqnarray}
Z_\ell = \frac{1}{S^{(0)}_\ell(v\to 1)}.
\end{eqnarray}

\begin{figure}
\begin{center}
\includegraphics[width=.45\textwidth]{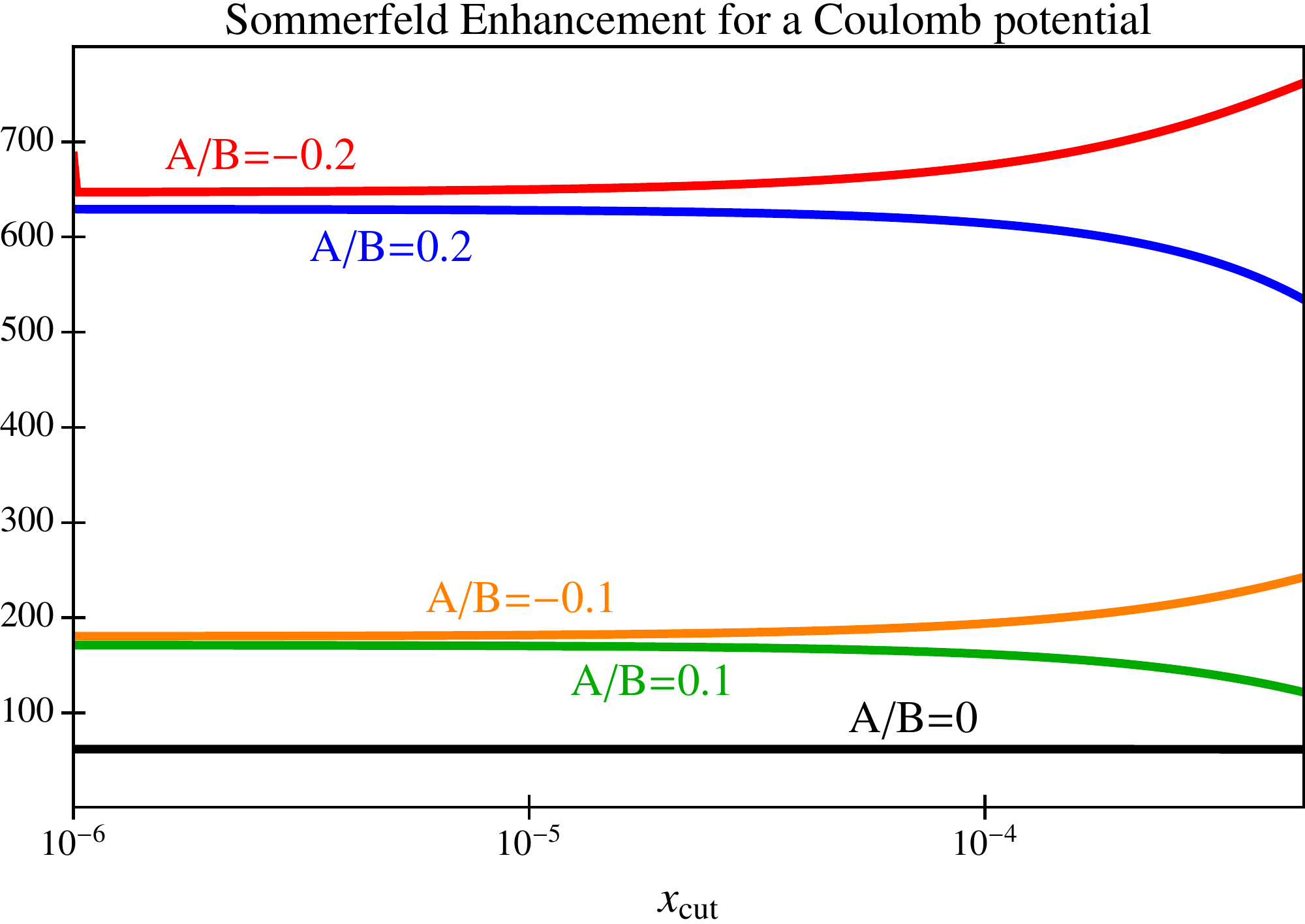}\;
\includegraphics[width=.45\textwidth]{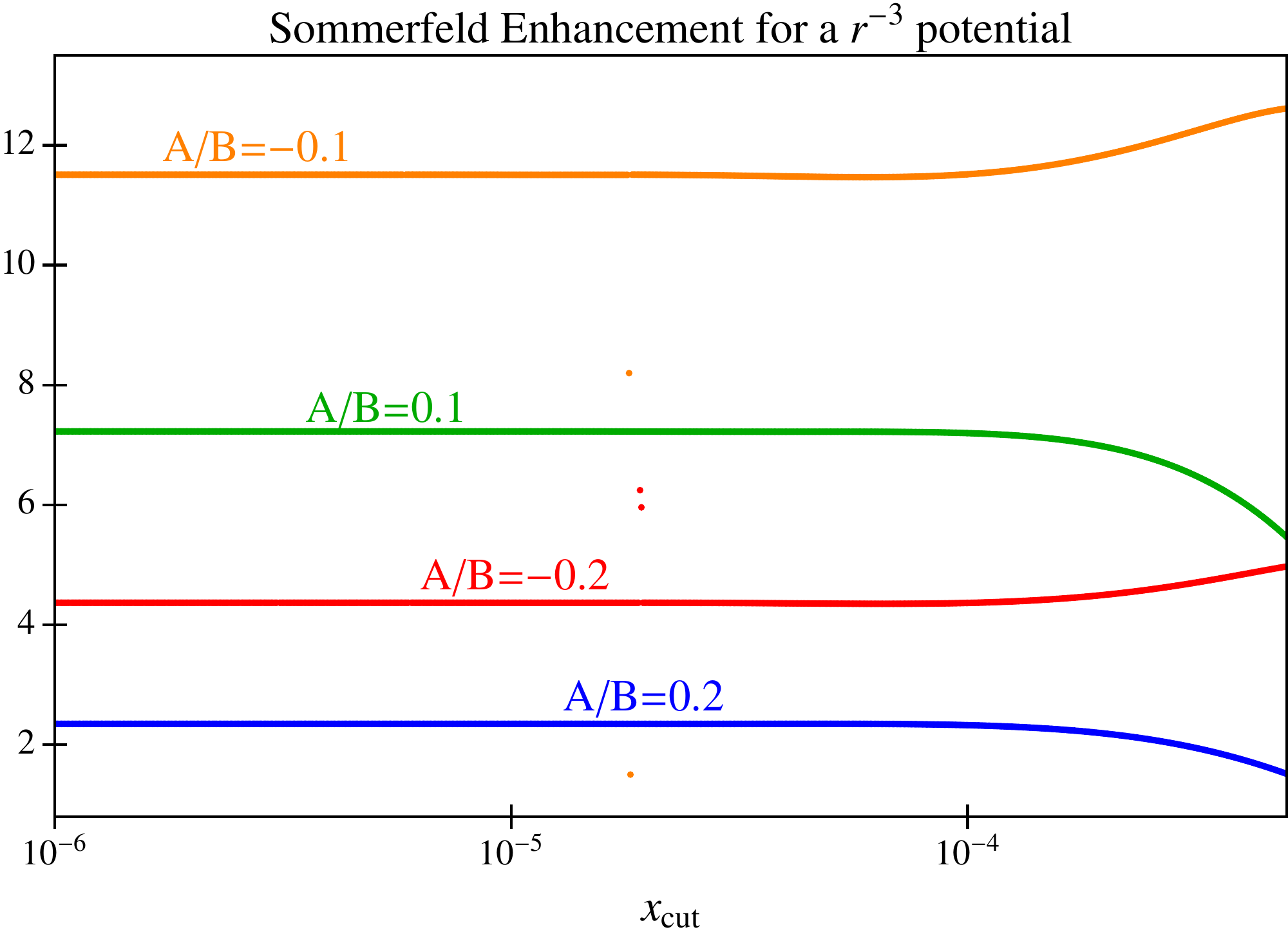}
\end{center}
\caption{Cutoff-dependence of $s$-wave Sommerfeld enhancement using the procedure described in the text. Low energy data is encoded by the ratio $A/B$ in (\ref{eq:schro:sol:gen:x:gtr:xcut}). We take relative velocity $v=10^{-3}$.
Deviations from flatness reflect a breakdown of the $x_\text{cut}\ll 1$ approximation.
\textsc{left:} 
    Coulomb potential with $\alpha/v=e^2/4\pi v=10$.
The unique phase ($A/B=0$) given by a \textsc{qed}-like \textsc{uv} completion is indicated by the black line.
\textsc{right:} $r^{-3}$ potential with $\tilde \alpha = 2M^2 v \alpha/f^2 = 10^{-3}$, for $\alpha$ defined in (\ref{eq:def:alpha}).
}
\label{r3:xcut:dep}
\end{figure}

\subsection{Comparison to Coulomb potential}

We now verify that the above procedure matches the usual result for the non-singular Coulomb potential, $V(r)=-\alpha/r$.
The wavefunction in the region $x>x_\text{cut}$ is 
\begin{eqnarray}
\Phi_{p,\ell}(x>x_\text{cut}) = 
    Ax^{1/2} J_{2\ell+1} \left(2 \sqrt{\frac{x\alpha}{v}}\right)
    + Bx^{1/2}Y_{2\ell+1}\left(2\sqrt{\frac{x\alpha}{v}}\right).
\end{eqnarray}
One can check that the Sommerfeld enhancement is indeed independent of the choice $x_\text{cut}$.
%
%
%
For different choices of $A/B$, one can obtain different Sommerfeld enhancements, as seen by the different lines 
on the left plot of Fig.~\ref{r3:xcut:dep}. 
Of these, one line (black) corresponds to the analytical formulae found in the literature \cite{Iengo:2009ni}; this corresponds to picking a scattering phase
that is consistent with a  relativistic completion that includes a massless boson. In other words, this is the choice that is consistent with a theory where the non-relativistic Coulomb potential is completed by a relativistic  field theory resembling \textsc{qed}. Other choices correspond to theories whose non-relativistic limit is Coulomb but whose local interactions differ from pure \textsc{qed}.

%
%
%

\section{Numerical results}
\label{sec:numerical}

The general \textsc{dm} potential considered here does not generally conserve orbital angular momentum $\vec{L}^{2}$ so that a coupled channel analysis between different $\ell$-wave annihilation modes is required. 
This implies that the $g_3$ contribution in (\ref{eq:potentialCP}) can still be relevant for Sommerfeld enhancement via $\Delta\ell=2$ transitions even though it averages to zero for $\ell=0$ states \cite{Bedaque:2009ri}. 
This is contrary to the common belief that Sommerfeld enhancement is relevant only for $s$-wave annihilations due to the centrifugal barrier.
%
%
%
%
For some states $\vec{L}^{2}$ is a well-defined quantum number once the total angular momentum $J$, the total spin $S$ and parity $P=\pm$ are specified. In these cases the calculation of the boost factor reduces to a standard single-channel Schr\"odinger problem as discussed above. Table~\ref{table:TableStates} shows the quantum numbers for fermionic \textsc{dm} for low total angular momenta. Among the $\ell=0$ states,  ($J=0$, $S=0$, $P=-$) gives a single channel problem with arbitrary potential $V_0(r)$, whereas ($J=1$, $S=1$, $P=-$) requires a coupled channel analysis between $\ell=0$ and $\ell=2$.

\begin{table}[h]
\begin{equation*}
\begin{array}{cccc}
\hline
J & S & P & \ell \\
\hline
0 & 0 & - & 0\\
 0 & 1 & + & 1 \\
 1 & 0 & +  & 1 \\
 1 & 1 & +  &  1  \\
 1 & 1 & -  &  0,\,2 \\
 \hline
\end{array}
\end{equation*} 
\caption{Low total angular momentum, $J$, \textsc{dm} scattering states labelled by spin, $S$, parity, $P$, and orbital angular momentum $\ell$. $J$, $S$, and $P$ are conserved by the Hamiltonian and are used to label states.}
\label{table:TableStates}
  \end{table}

Assuming parity conservation, the effective potential $V_{\text{eff}}=\langle\text{out}|V(r)|\text{in}\rangle+\ell(\ell+1)/(2Mr^2)$ for each channel is obtained by sandwiching (\ref{eq:potential:parity}) with the centrifugal term between the appropriate $\ket{\,J\; S \; P\,}$ states,
%
\begin{eqnarray}
\ket{\;0\quad 0 \quad -}  &\rightarrow& V_{\text{eff}}=\left(\tilde{g}_1(r)-\frac 34 \tilde{g}_2(r)\right) \frac{1}{4\pi r}\\
\ket{\;0\quad 1 \quad +}&\rightarrow& V_{\text{eff}}=\frac{1}{Mr^2}+\left(\tilde{g}_1(r)+\frac 14 \tilde{g}_2(r)-\frac{\tilde{g}_3(r)}{2\Lambda^2 r^2}-\frac{2\tilde{g}_7(r)}{M\Lambda r^2} \right) \frac{1}{4\pi r}\\
\ket{\; 1\quad 0\quad +} &\rightarrow& V_{\text{eff}}= \frac{1}{Mr^2} +\left(\tilde{g}_1(r)-\frac 34 \tilde{g}_2(r)\right)\frac{1}{4\pi r}\\
\ket{\;1 \quad 1 \quad +} &\rightarrow& V_{\text{eff}}=\frac{1}{Mr^2}+\left(\tilde{g}_1(r)+\frac 14 \tilde{g}_2(r)+\frac{\tilde{g}_3(r)}{4\Lambda^2 r^2} -\frac{\tilde{g}_7(r)}{M\Lambda r^2}  \right) \frac{1}{4\pi r}\\
\ket{\; 1 \quad 1 \quad-} &\rightarrow& V_{\text{eff}}=
\frac{1}{Mr^2}\left(\begin{matrix} 0 & 0\\ 0 & 3\end{matrix}\right)+
\left(\begin{matrix} \tilde{g}_1(r)+\frac{\tilde{g}_2(r)}{4} & \frac{\tilde{g}_3(r)}{2\sqrt{2}\Lambda^2 r^2}\\ \frac{\tilde{g}_3(r)}{2\sqrt{2}\Lambda^2 r^2} & \tilde{g}_1(r)+\frac{\tilde{g}_2(r)}{4}-\frac{\tilde{g}_3(r)}{4\Lambda^2 r^2}-\frac{3\tilde{g}_7(r)}{M\Lambda r^2} \end{matrix}\right) \frac{1}{4\pi r}
\label{eq:coupled:channel:Veff}
\end{eqnarray}
where the $\ell=0$ and $\ell=2$ channels are coupled in (\ref{eq:coupled:channel:Veff}). 
If the $\tilde{g}_i$ are constant, then at leading order these channels are effectively non-singular and Coulomb-like.  However, if $\tilde{g}_1+\tilde{g}_2/4=0$, such as for pseudo-scalar exchange, then some of these channels are dominated by the singular $V\sim 1/r^3$ term.  
Moreover, one can also consider scenarios---for example, the exchange of multiple light particles \cite{Ferrer:1998rw,Hsu:1992tg,Feinberg:1989ps, Dobrescu:2006au}---in which $\tilde{g}_{1,2}\sim 1/r^3$ so that even the $\tilde{g}_1$ and $\tilde{g}_2$ terms are singular with $\ell=0$.
Thus one may in principle generate a singular potential for any partial wave. For simplicity, we shall consider a simple $1/r^3$ potential for both $\ell=0$ and $\ell=1$. The coupled channel in (\ref{eq:coupled:channel:Veff}), however, requires a more careful analysis that we leave for future work.

\begin{figure}
\begin{center}
\includegraphics[width=.45\textwidth]{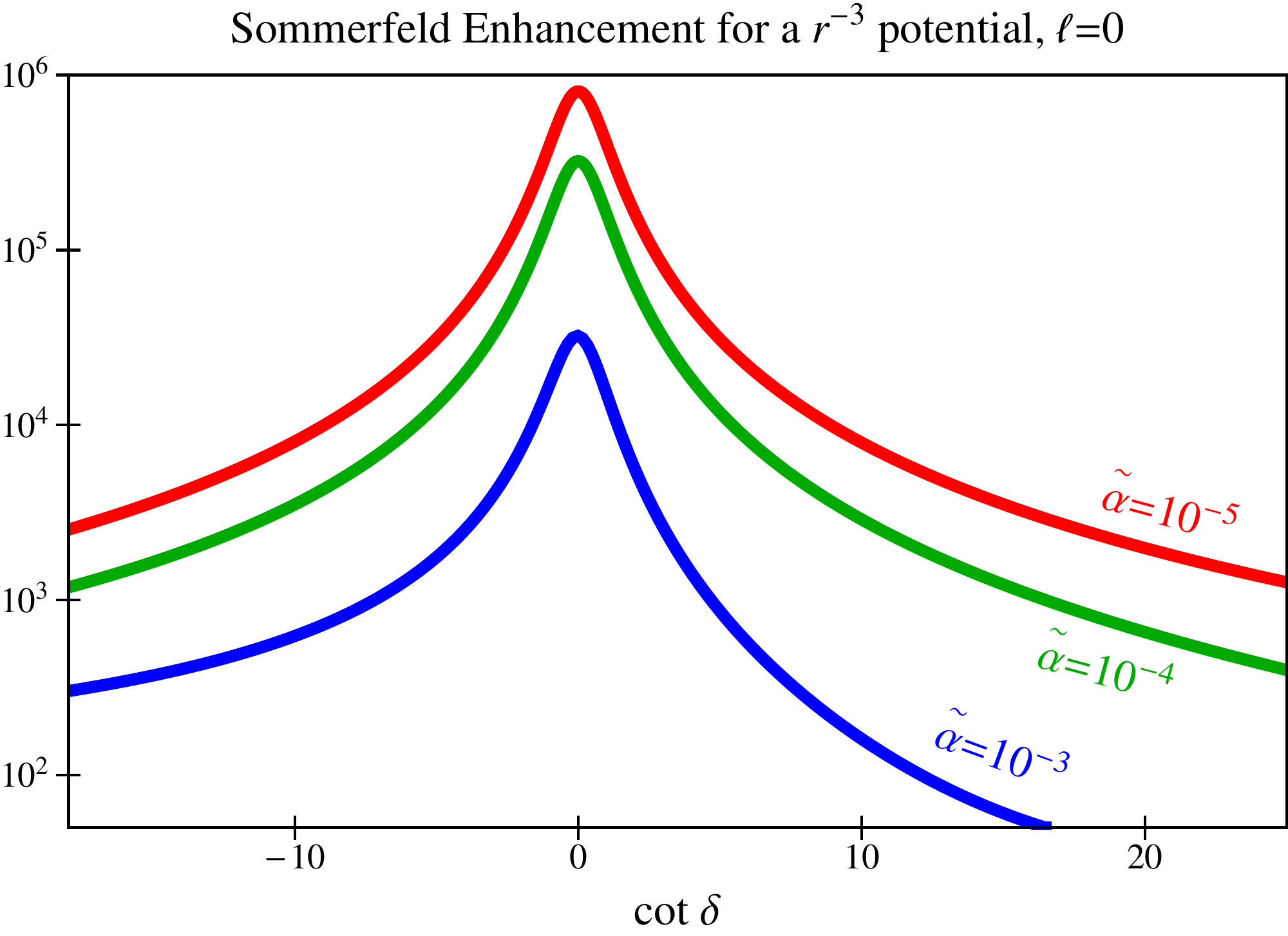}\,
\includegraphics[width=.45\textwidth]{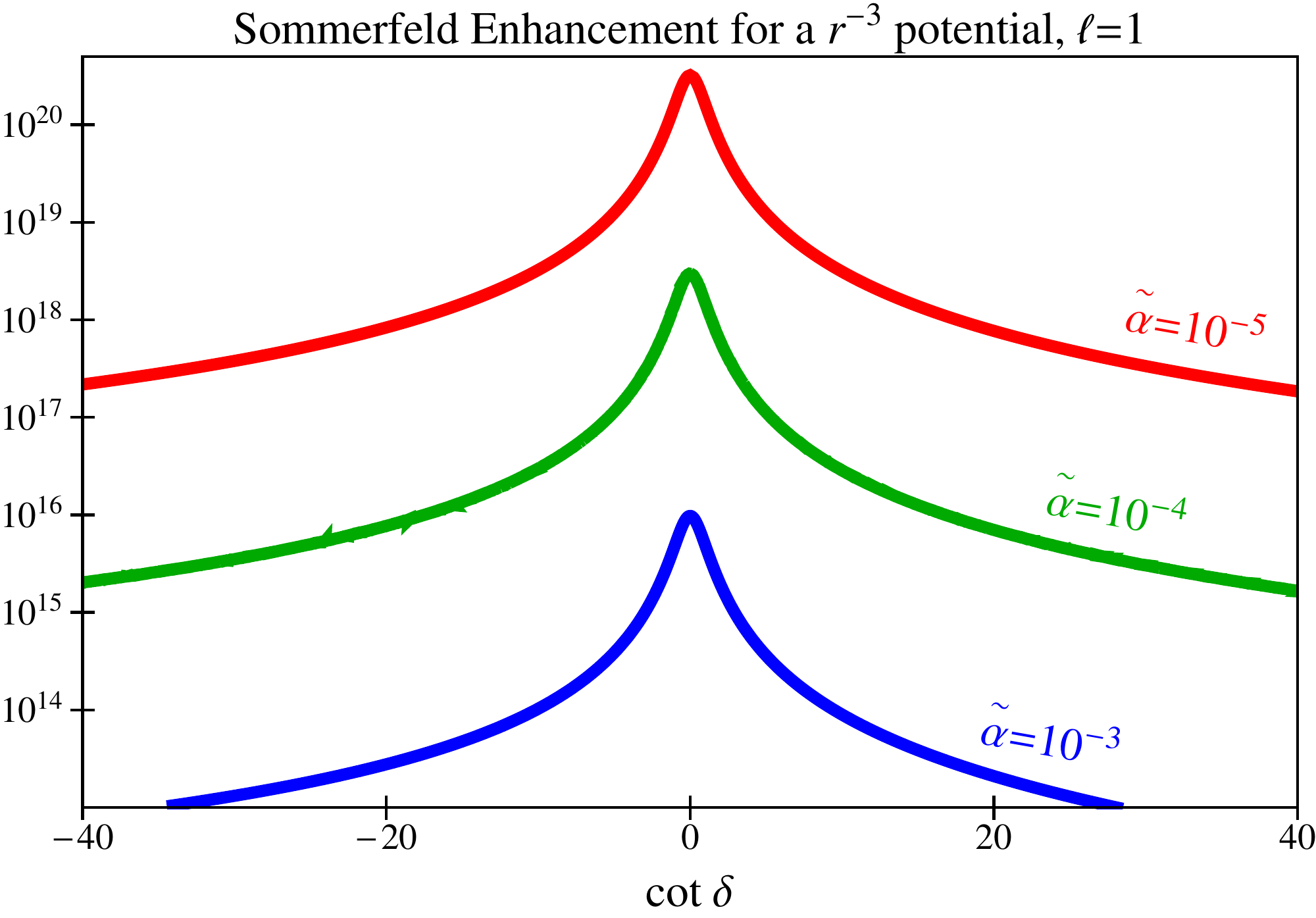}
\end{center}
\caption{
Sommerfeld enhancement for a singular $r^{-3}$ potential and orbital angular momentum $\ell=0$ (\textsc{left}) and $\ell=1$ (\textsc{right}) for relative velocity $v=10^{-3}$ and various values of  $\tilde \alpha = 2M^2 v \alpha/f^2$, with $\alpha$ defined in (\ref{eq:def:alpha}). } 
\label{fig:Somcot}
\end{figure}

In Fig.~\ref{fig:Somcot} we plot the Sommerfeld enhancement for a potential
\begin{align}
V(r)=-\frac{\alpha}{f^2r^3}
\label{eq:def:alpha}
\end{align}
as a function of the \textsc{ir} observable $\cot\delta$ for $\ell={0,1}$. When comparing these, note that the $\ell=1$ cross section has an additional factor of $v^2$ relative to $\ell=0$.  The resonance is located at $\cot \delta=0$ because this is where the cross section is maximal. These plots can be used to give an upper bound on Sommerfeld enhancement for various couplings.  Note that while it is true that the resonance is larger for smaller couplings, it requires more tuning from the \textsc{uv} to reach the resonance for a smaller coupling.  Moreover, while $\cot\delta$ contains data about \textsc{uv} physics, it also depends on the \textsc{ir} coupling in such a way that reducing the coupling would not increase the Sommerfeld unless one simultaneously increases the height of the square well potential $V_0$.

\begin{figure}
\begin{center}
\includegraphics[width=.45\textwidth]{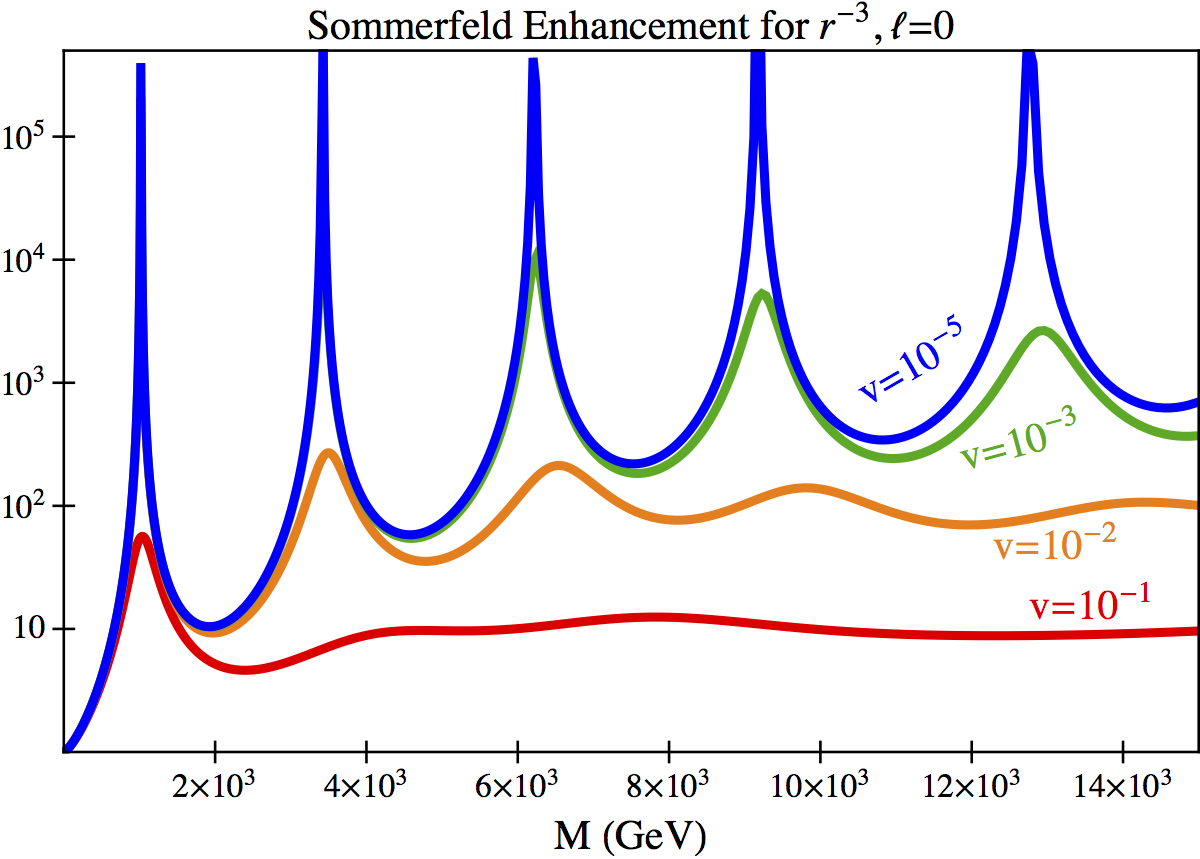}\,
\includegraphics[width=.45\textwidth]{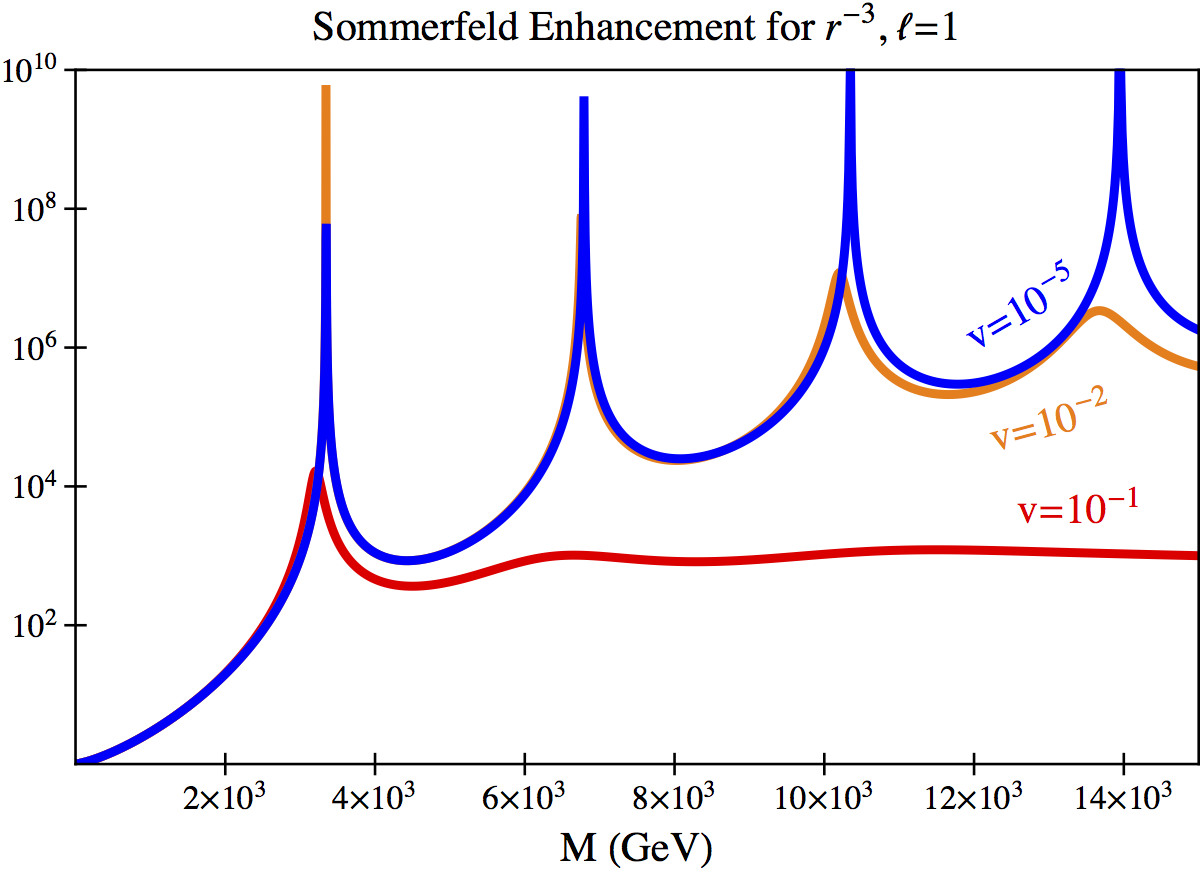}
\end{center}
\caption{
Resonances in Sommerfeld enhancement for a singular $r^{-3}$ potential and orbital angular momentum $\ell=0$ (\textsc{left}) and $\ell=1$ (\textsc{right}) for a range of relative velocities and $\alpha/f^2$ = $\text{TeV}^{-2}$ with  $\alpha$ defined in (\ref{eq:def:alpha}).
The large enhancements can be understood from the box approximation, see the Appendix~\ref{app:box}. For simplicity the height of the regulated potential is fixed by continuity with the long range piece.} 
\label{fig:NDA}
\end{figure}

Fig.~\ref{fig:NDA} presents an exploration of these resonances as a function of the  dark matter reduced mass. As described in the procedure above, the physical Sommerfeld enhancement for a singular potential requires information from an \textsc{ir} observable such as the scattering phase $\delta$. 
As a reasonable estimate for natural \textsc{uv} models, we regulate the theory at a length scale $r_0$ where the non-relativistic description breaks down, $V(r_0)=M$. We then fix the height of the cutoff by continuity with the singular long-range part, $V_0=V(r_0)=M$.  Notice that for a $V(r)=-\alpha/(f^2r^3)$ potential with $f=1$ TeV,  the dark matter mass necessary to reach a significant enhancement is about $1$ TeV.  If the dark matter mass is sufficiently large one may also need to consider the $\ell=1$ contribution. This appears to contradict the common belief that $\ell>0$ enhancement is too velocity suppressed to be relevant.

\section{Phenomenology}


While the collisionless cold \textsc{dm} paradigm successfully accounts for the large scale structure of the universe, it faces tension at smaller scales where $N$-body simulations present some discrepancies with observations. In particular, dwarf galaxies show flat core \textsc{dm} densities profiles in the central part of the halos, whereas collisionless cold \textsc{dm} predicts cusp-like profiles \cite{Oh:2010ea, deNaray:2011hy,Flores:1994gz, Simon:2004sr}. While this discrepancy may be due to unaccounted baryonic physics \cite{blumenthal1986contraction, Gnedin:2004cx, Tissera:2009cm}, it may alternately be taken as a motivation for dark matter self-interactions \cite{Vogelsberger:2012ku, Peter:2012jh, Rocha:2012jg}.
A related astrophysical motivation for self interactions is the ``too big to fail problem,'' in which the brightest observed dwarf spheroidal satellites in the Milky Way appear to be incompatible with the central densities of subhalos predicted by collisionless \textsc{dm}\cite{Sawala:2010zw, BoylanKolchin:2011de, BoylanKolchin:2011dk}.
A third suggestion for self interactions is the ``missing satellites problem'';  collisionless \textsc{dm} predictions for the number the satellite galaxies expected in the Milky Way appears to disagree with observations \cite{Moore:1999nt, Klypin:1999uc}.
See, e.g.~\cite{Buckley:2009in, Tulin:2013teo} and references therein for critical discussions.


To solve the core vs.~cusp problem, the dark matter self interaction must have a sufficiently large cross section, $\sigma/m_\chi\sim 0.1-10 \text{ cm}^2/\text{g}$, for velocities typical of dwarf galaxies, $v\sim 10^{-5}$, while having a smaller cross section for galaxy cluster velocities, $v\sim 10^{-3}$, where collisionless \textsc{dm} results are in good agreement. 
There are additional upper bounds on the cross section coming from astrophysical observations sensitive to the velocities characteristic of galaxy clusters \cite{Buckley:2009in, Feng:2009hw}. One of the most stringent bounds, for example, comes from the ellipticity of galaxy clusters \cite{MiraldaEscude:2000qt, Feng:2009hw, Oguri:2010ik}. The most recent simulations have softened this bound to $\sigma/m_\chi = 0.1 \text{ cm}^2/\text{g}$ \cite{Rocha:2012jg, Peter:2012jh}. Further, the cosmic microwave background (\textsc{cmb}) sets an upper bound on Sommerfeld enhancement from the effect of \textsc{dm} annihilation after recombination \cite{Finkbeiner:2010sm, Hannestad:2010zt, Hisano:2011dc}. 
Though a constant cross section $\sigma/m_\chi \lesssim 0.5 \text{ cm}^2/\text{g}$ may account for these effects, this velocity dependence is also suggestive of a Sommerfeld enhanced cross section \cite{Loeb:2010gj}. 
We leave a more thorough investigation of the astrophysical and cosmological bounds on the enhancement of singular potentials for future work.

\begin{figure}
\begin{tabular}{ll}
\includegraphics[width=.43\textwidth]{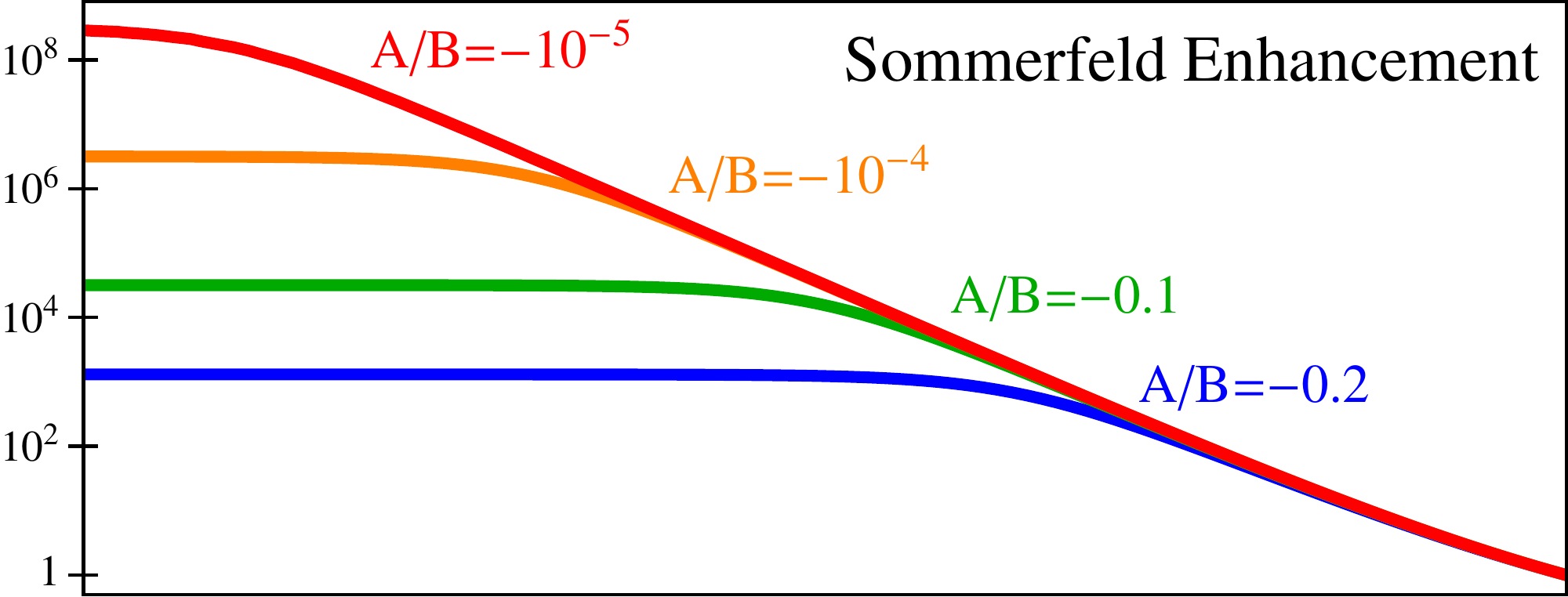} &
\multirow{2}{*}[6em]{
    \includegraphics[width=.55\textwidth]{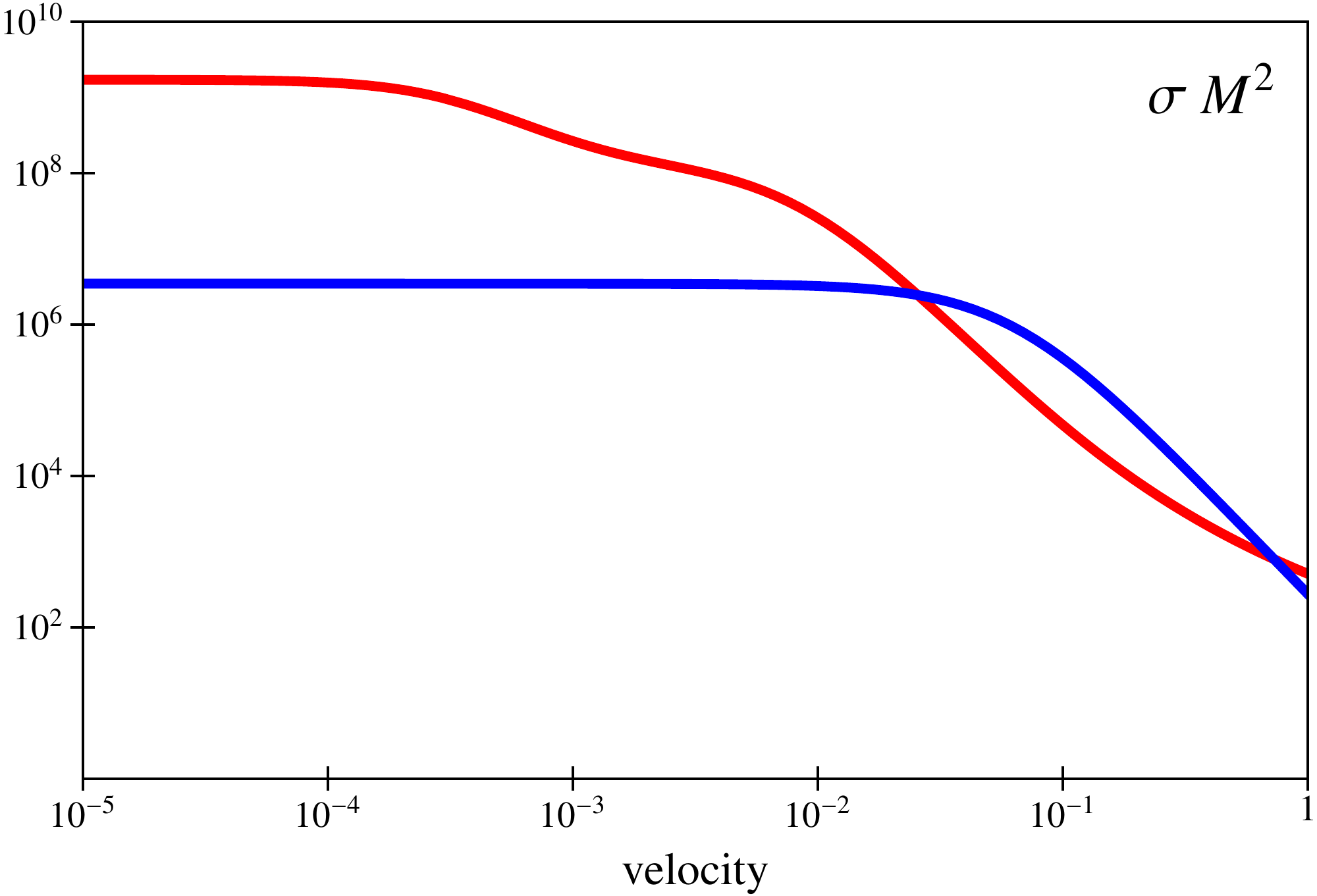}
}\\

\includegraphics[width=.43\textwidth]{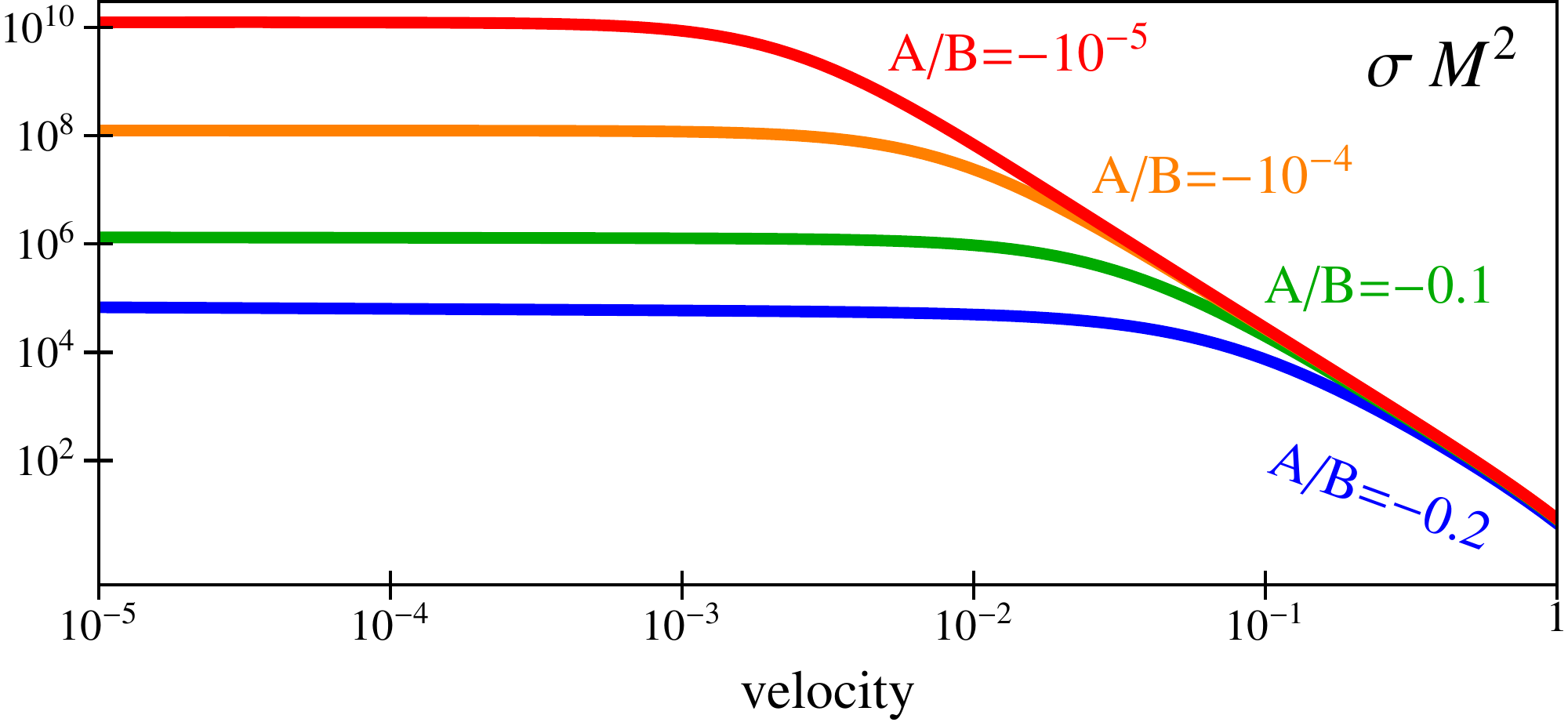}

\end{tabular} 
%
\caption{Core vs.~cusp problem. 
\textsc{left}: Sommerfeld enhancement (upper) and scattering cross section (lower) as a function of relative velocity for a range of low energy parameters $A/B$ as discussed below (\ref{eq:schro:sol:gen:x:gtr:xcut}) and $2\alpha M^2/f^2 = 1$.
%
%
\textsc{right}: Total dark matter cross section as a function of velocity. Red: velocity dependent with $2\alpha M^2/f^2 = 34$, $M$ = 5.8 TeV, $A/B = -10^{-3}$, and an additional short distance interaction, $M^2 \sigma_\text{short}$ = 500. Blue: velocity independent cross section with no new short range interaction and $M=$ TeV, $2\alpha M^2/f^2 = 0.1$, $A/B = -6\times 10^{-4}$.
} \label{fig:corecusp}
\end{figure}

\begin{figure}
\begin{center}
\includegraphics[width=.45\textwidth]{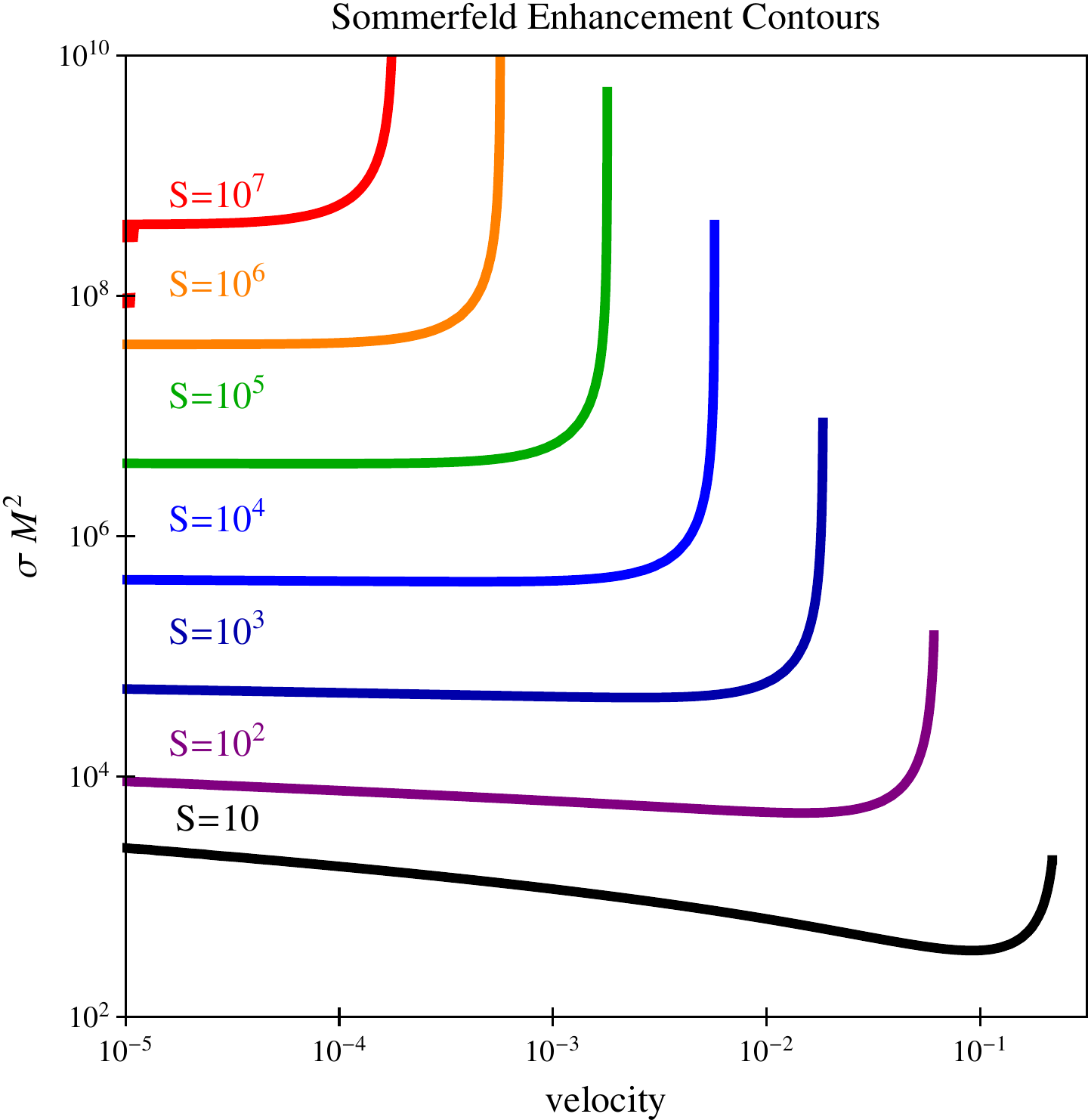}
\end{center}
\caption{Contours of Sommerfeld enhancement from the singular potential (\ref{eq:pheno:singular:potential}) with $2\alpha M^2/f^2=1$ as a function of the \textsc{dm} velocity and elastic cross section $\sigma$.} 
\label{fig:contour}
\end{figure}

As an example for how to apply Sommerfeld enhancement to address the dwarf galaxy scale astrophysical puzzles while simultaneously avoiding the bounds from galaxy cluster scale observations, we consider dark matter self interactions mediated by a light force carrier that generates a singular potential,
\begin{align}
V(r) = \frac{-\alpha}{f^2}\frac{1}{r^3}.
\label{eq:pheno:singular:potential}
\end{align}
The left side of Fig.~\ref{fig:corecusp} shows the Sommerfeld enhancement (upper) and the total cross section (lower) from such a model with a choice of parameters near the resonance. Observe that even for very small $A/B$, that is small $\cot \delta$ or large scattering phase, the cross section is saturated between the characteristic galaxy cluster velocities $v\sim 10^{-5}$ and dwarf galaxy velocities $v\sim 10^{-3}$. For $A/B\sim 10^{-5}$, as indicated by the red line in the lower figure, this saturates to $\sigma/m_\chi \sim 10^{-2} \text{ cm}^2/\text{g}$ for $m_\chi \sim \text{TeV}$. 
%
This saturation occurs over the range of velocities where we would like a stronger velocity-dependence to avoid cluster scale bounds.
In order to do this, we assume the existence of a short range interaction that contributes to the elastic scattering process with cross section $\sigma_\text{short}^{(0)}$. The long range mediators Sommerfeld enhance this cross section by the factor shown in the upper plot; observe that this enhancement decreases exponentially as one increases from dwarf galaxy velocities to galaxy cluster velocities. The total cross section is roughly (ignoring cross terms for simplicity),
\begin{align}
\sigma_\text{tot}(v)\sim\sigma_\text{elast}(v)+S(v)\sigma^{(0)}_\text{short}.
\end{align}
Since the enhancement factors can be fairly large, the additional short range interaction can be weakly coupled, e.g.\ $\sigma_\text{short}^{(0)} M^2 \sim 10^4$ so that $\sigma_0 \sim 10^6 \text{ pb}$ for a TeV scale dark matter particle. The right side of Fig.~\ref{fig:corecusp} compares the velocity-dependence of this type of solution to another solution without enhanced short range physics.
Fig.~\ref{fig:contour} shows contours of Sommerfeld enhancement as a function of velocity and elastic cross section, combining the data from the left-hand side of Fig.~\ref{fig:corecusp}.



Finally, we remark on the use of Sommerfeld enhancement for generating indirect signals of dark matter through positrons and gamma rays \cite{ArkaniHamed:2008qn, Hisano:2004ds}. The excess of cosmic positrons observed by \textsc{pamela} \cite{Adriani:2008zr} and later confirmed by \textsc{Fermi} \cite{FermiLAT:2011ab}  and \textsc{ams-02} \cite{Aguilar:2013qda} is a potential signal for dark matter annihilation. Since the cross section required to produce these signals is much larger than the required cross section for thermal relics, \textsc{dm} models that realize the positron excess typically require large Sommerfeld enhancements \cite{Bergstrom:2009fa}. A study for non-singular dark sectors with Yukawa interactions was performed in \cite{Feng:2010zp, Feng:2009hw}; an investigation of how these bounds change for singular potentials is left for future work.


%
A recent speculative signal of indirect \textsc{dm} detection is the 135 GeV line in the \textsc{Fermi} gamma ray spectrum \cite{Weniger:2012tx, Su:2012ft, Rajaraman:2012db, Bloom:2013mwa, Ackermann:2012qk}. Indeed, gamma ray signatures were the original motivation for investigating Sommerfeld enhancement in dark matter \cite{Hisano:2004ds}.  The cross section required for the line is about $10^{-27}\text{ cm}^3/\text{s}$ which generically points toward a large boost factor, $S\approx 10^{4}$.
It is possible to get such a large enhancement with a singular potential $V(r)=-\alpha/(fr^3)$, but since the dark matter mass must be 135 GeV this requires a low scale $f\approx 100 \text{ GeV}$ to avoid tuning in the \textsc{uv}.  
Dark matter models can generate such a feature, though these typically generate an unobserved continuum contribution to the spectrum \cite{Cohen:2012me}. Ways around difficulty were explored in \cite{Fan:2012gr, Jackson:2009kg, Jackson:2013tca, Jackson:2013pjq, Goodman:2010qn}.
%

%
%
%
%


%

\section{Conclusion}

We have presented the effective non-relativistic theory of self-interacting dark matter
parameterized to leading order in the relative velocity, $v$, and the exchanged momentum, $q/\Lambda$.
The resulting potentials generically include singular terms which must be regulated and renormalized so that the resulting predictions are cutoff independent.
We have shown how this effective theory can be applied to calculate the Sommerfeld enhancement generated by singular potentials. 

Using a simple toy model with a $1/r^3$ potential, we have found that on resonance one can generate enhancements as large as $S \sim 10^6$ at velocities on the order of $v\sim 10^{-3}$. This opens up promising directions for the astrophysical phenomenology of general self-interacting dark matter models. For example, extant astrophysical puzzles such as the core vs.\ cusp problem can be addressed with this velocity-dependent enhancement. A more thorough investigation and implications for specific \textsc{uv} models of these bounds is left for future work.

\section*{Acknowledgements}
\textsc{b.b.}~is supported in part by the \textsc{miur-firb}
grant \smaller[0.85]{RBFR12H1MW}, 
by the \textsc{erc} Advanced Grant No.~267985, ``Electroweak Symmetry Breaking, Flavour and Dark Matter: One Solution for Three Mysteries'' (DaMeSyFla), and by the \textsc{nsf} through grant \smaller[0.85]{PHY11-25915}.
\textsc{m.c.}~and \textsc{p.t.}~ are supported by the \textsc{nsf} through grant \smaller[0.85]{PHY-0757868}.
The authors thank Jonathan Feng, Peter Lepage, Maxim Perelstein, Bibhushan Shakya, Tim Tait, and Itay Yavin for helpful discussions.
%
%
\textsc{b.b.}~thanks the \textsc{kitp} at \textsc{ucsb}
for the hospitality during the workshop ``Exploring TeV Scale New Physics with \textsc{lhc} Data'' where part of this work was completed. \textsc{b.b.}~thanks Cornell University for its hospitality during the course of this work. 


\appendix

\section{$CP$-preserving  potential}
\label{app:list}

In Section~\ref{sec:effectivesect}, we presented a list of $P$- and $T$-preserving operators in the non-relativistic potential for \textsc{dm} self-interactions.
In this appendix we present the additional terms in the effective potential that are generated when parity invariance is relaxed.
In addition to $\mathcal{O}_{1,2,3,4,7,8}$, the four operators $\mathcal{O}_{9,10,11,12}$ in (\ref{eq:O9})--(\ref{eq:12}) preserve $CP$ but break parity.  For simplicity we consider only the case of self-conjugate \textsc{dm} so that $\mathcal{O}_{10,12}$ are forbidden. 

The $\mathcal{O}_9$ term contains no $\vec{v}_\perp$ factors and the corresponding potential is
\begin{equation}
V_9=\frac{\tilde{g}_9(r)}{4\pi r^3\Lambda } (\vec{s}_1\times\vec{s}_2)\vec{r}
\end{equation}
where $\tilde{g}_9(r)$ is defined analogously to (\ref{dofr}). 

In order to determine $V_{11}$ we need the Fourier transform  of the propagator along the direction tranverse to the exchanged momentum
\begin{align}
 & \int \frac{d^3 \vec{q}}{(2\pi)^3}e^{i\vec{q}\cdot \vec{r}}\left[\delta^{ij}-\frac{\vec{q}_i \vec{q}_j}{\vec{q}^2}\right]\frac{1}{(\vec{q}^2+\mu^2)} 
=& \frac{e^{-\mu r}}{4\pi r}\left[\frac{2}{3}\delta^{ij}+\frac{1}{\mu^2 r^2}\left(3\hat{r}^i \hat{r}^j-\delta^{ij}\right)\left(e^{\mu r}-1-\mu r-\frac{\mu^2 r^2}{3}\right)\right]\,.\nonumber
\end{align}
Contracting this expression  with $(\vec{s}_1-\vec{s}_2)^i$ and $\vec{v}^j$ gives $V_{11}$. 
Since the final result is quite involved, we focus on two interesting limits. At distances smaller than the mediator Compton wavelength, $\Lambda^{-1}\ll r\ll \mu^{-1}$, the expression greatly simplifies because
\begin{equation}
\lim_{\mu\rightarrow 0}\int \frac{d^3 \vec{q}}{(2\pi)^3}e^{i\vec{q}\cdot \vec{r}}\left[\delta^{ij}-\frac{\vec{q}_i \vec{q}_j}{\vec{q}^2}\right]\frac{1}{(\vec{q}^2+\mu^2)}= \frac{1}{8\pi r}\left(\delta^{ij}+\hat{r}^i \hat{r}^j\right),
\end{equation}
and hence 
\begin{equation}
V_{11}=\frac{1}{8\pi r}\left[(\vec{s_1}-\vec{s}_2)\cdot \vec{v}+(\vec{s_1}-\vec{s}_2)\cdot \hat{r}  (\hat{r}\cdot\vec{v})\right]\,.
\end{equation}
On the other hand, at scales where the mediator mass is important, $r\gg \mu^{-1}$, we have
\begin{equation}
V_{11}=\frac{1}{4\pi r^3 m^2 } \left[3(\vec{s_1}-\vec{s}_2)\cdot \hat{r}\,\,\,(\hat{r}\cdot\vec{v})-(\vec{s_1}-\vec{s}_2)\cdot \vec{v}\right]\,.
\end{equation}
where $m^2=\int d\mu^2 \rho(\mu^2)/\mu^2$.

We stress that the ordering of the various operators in the non-static part of the potential is generically important since $\vec{p}=m_\chi \vec{v}/2$ is the conjugate coordinate associated with the relative distance, $[\vec{r}^i,\vec{p}^j]=i\delta^{ij}$.

\section{Sommerfeld enhancement for non-singular potentials}
\label{app:origin:of:sommerfeld}

Let us first briefly review the general method to obtain the Sommerfled enhancement \cite{Iengo:2009xf,Iengo:2009ni}. Consider two particles of mass $m_\chi$ and center-of-mass momentum $\vec{p}$. The $\ell$-wave amplitude $A_\ell(\vec{p})$ for the annihilation of these two particles under an attractive central potential $V(r)$ can be expressed as a function of a bare amplitude $A_{0,\ell}(\vec{q})= a_{0,\ell} q^\ell$ and a wavefunction $\phi_\vec{p}(\vec{r})$,
\begin{align}
	A_\ell(\vec{p}) &= 	\int d\vec{r}\, \phi^*_\vec{p}(\vec{r})
						\int d\vec{q}\, e^{i\vec{q}\cdot\vec{r}} A_{0,l}(\vec{q})
						\label{eq:amp}.
\end{align}
The wavefunction $\phi_{\vec{p}}(\vec{r})$ satisfies the Schr\"odinger equation,

\begin{align}
	\left(-\frac{1}{2M}\partial^2+V(r)-\frac{p^2}{2M}  \right)\phi_{\vec{p}}(\vec{r}) & =0,\label{eq:schro}
\end{align}
where $M=m_{\chi}/2$ is the reduced mass and $p=M v$ is the non-relativistic momentum.  In general, the potential $V(r)$ can be matrix valued in the space of partial waves, in which case the Schr\"odinger equation is then a system of coupled differential equations.  To solve this equation we decompose the wavefunction $\phi_{\vec{p}}(\vec{r})$ in partial waves  
\begin{align}
	\phi_{\vec{p}}(\vec{r}) &= \frac{(2\pi)^{3/2}}{4\pi p}\sum_{\ell} 
							(2\ell+1)e^{i\delta_\ell}R_{p,\ell}(r)
							P_\ell(\hat{\vec{p}}\cdot\hat{\vec{r}})
							\label{eq:partial}
\end{align}
such that the radial part, $R_{p,\ell}(r)$, satisfies
\begin{align}
	\frac{-1}{2M}
	\left(	\frac{d^2R_{p,\ell}}{dr^2}
			+\frac{2}{r}\frac{dR_{p,\ell}}{dr}
			-\frac{\ell(\ell+1)}{r^2}R_{p,\ell}
	\right)
	- \left( \frac{p^2}{2M}-V(r) \right) R_{p,\ell}
	& = 0 \label{eq:Rschro}
\end{align}
with the completeness relation
\begin{align}
	\int_{0}^{\infty}dp R_{p,\ell}(r)R_{p,\ell}(r')
	& =
	\frac{\delta(r-r')}{r^2}.
	\label{eq:compl}
\end{align}
Plugging the partial wave decomposition (\ref{eq:partial}) into (\ref{eq:amp}) along with $\phi^{0}_{\vec{p}}(\vec{r})=e^{i\vec{p}\cdot\vec{r}}$ gives
\begin{align}
	A_l(p) &= \frac{1}{p} \int_{0}^{\infty} r^2dr 
				R_{p,\ell}(r)\int_{0}^{\infty} qdq R^0_{q,\ell}(r)A_{0,\ell}(q).
				\label{eq:Al}
\end{align}
From the free solution $R^{0}_{p,\ell}$ we know that
\begin{align}
	\frac{d^\ell}{dr^\ell} \, R^0_{q,\ell}(r=0)
	& = \sqrt{\frac{2}{\pi}}\frac{\ell!q^{\ell+1}}{(2\ell+1)!!}.
\end{align}
Applying the completeness relation (\ref{eq:Al}) gives
\begin{align}
	A_{\ell}(p,p') &= 
	\sqrt{ \frac{\pi}{2} }
	\frac{(2\ell+1)!!}{\ell!}
	\frac{1}{p}
	\frac{d^\ell}{dr^\ell}
	R_{p,\ell }(r=0) a_{0,\ell}
\end{align}
such that the Sommerfeld enhancement for a the $\ell^\text{th}$ partial wave is
\begin{align}
	S_{l} & = 
	\left|	\sqrt{\frac{\pi}{2}} 
			\frac{(2\ell+1)!!}{\ell!}
			\frac{1}{p^{\ell+1}}
			\frac{d^\ell}{dr^\ell}
			R_{p,\ell}(r=0)
	\right|^2 
	\label{eq:Smaster}
\end{align}
We thus see that the Sommerfeld enhancement is given by the solution of the Schr\"odinger equation at the origin. 

\subsection{Numerical algorithm}

Refs.~\cite{Iengo:2009ni, Iengo:2009xf} provide a method to numerically evaluate the enhancement factor $S$.  The completeness relation (\ref{eq:compl}) is valid at long distances,
\begin{align}
	\left.R_{p,\ell}(r)\right|_{r\rightarrow \infty}
	\rightarrow \sqrt{\frac{2}{\pi}}
	\frac{\sin(pr-\ell\pi/2+\delta_\ell)}{r}.
\end{align}
For simplicity, let us work with the dimensionless variable $x=pr$ and the rescaled wavefunction $\Phi_{p,\ell}(x)=\frac{xR_{p,\ell}(x)}{Np}$ where $N$ is an arbitrary normalization.  Using these variables, the Schr\"odinger equation takes the form
\begin{align}
	-\Phi_{p,\ell}(x)''
	+ \left( \mathcal V(x)+\frac{\ell(\ell+1)}{x^2}-1\right)
	\Phi_{p,\ell}(x)
	&=0 
	\label{eq:NSchro}
\end{align}
where $\mathcal V(x)=\frac{2M}{p^2}V(x/p)$ and we impose the initial conditions 
\begin{align}
	\lim_{x\rightarrow 0}
	\Phi_{p,\ell}(x)
	= x^{\ell+1}.  
	\label{eq:inic}
\end{align}
From (\ref{eq:NSchro}) and the fact that $\lim_{x\rightarrow \infty} \mathcal V(x) = 0$, it is clear that in the asymptotically far away region, 
\begin{align}
	\left. \Phi_\ell(x) \right|_{x\rightarrow \infty} 
	\rightarrow C\sin(x-\ell\pi/2+\delta_\ell)
\end{align}
Moreover, to satisfy the asymptotic normalization of $R_{p,\ell}(r)$, we need to fix the normalization $N=\sqrt{\frac{2}{\pi}}\frac{1}{C}$.  We can then use $R_{p,\ell}=N p \Phi_l/x$ in (\ref{eq:Smaster}) along with the initial condition to obtain
\begin{align}
	A_\ell(p)=\frac{(2\ell+1)!!}{C}p^\ell a_{0,\ell}
	= \frac{(2\ell+1)!!}{C}A_{0,\ell}(p)
\end{align}
so that the Sommerfeld factor is 
\begin{align}
	S = \left( \frac{(2\ell+1)!!}{C} \right)^2
\end{align}
We thus reduce the calculation of the Sommerfeld enhancement $S$ to the determination of $C$.  This is obtained by numerically solving (\ref{eq:NSchro}) with the initial condition (\ref{eq:inic}) and 
\begin{align}
	C^2 
	&=
	\left( \Phi_l(x)^2 + \Phi_l(x-\pi/2)^2 \right)
	\left|_{x\rightarrow\infty}\right. .
\end{align}

\subsection{Coulomb and Yukawa example}

\begin{figure}[t]
\centering
\includegraphics[width=.75\textwidth]{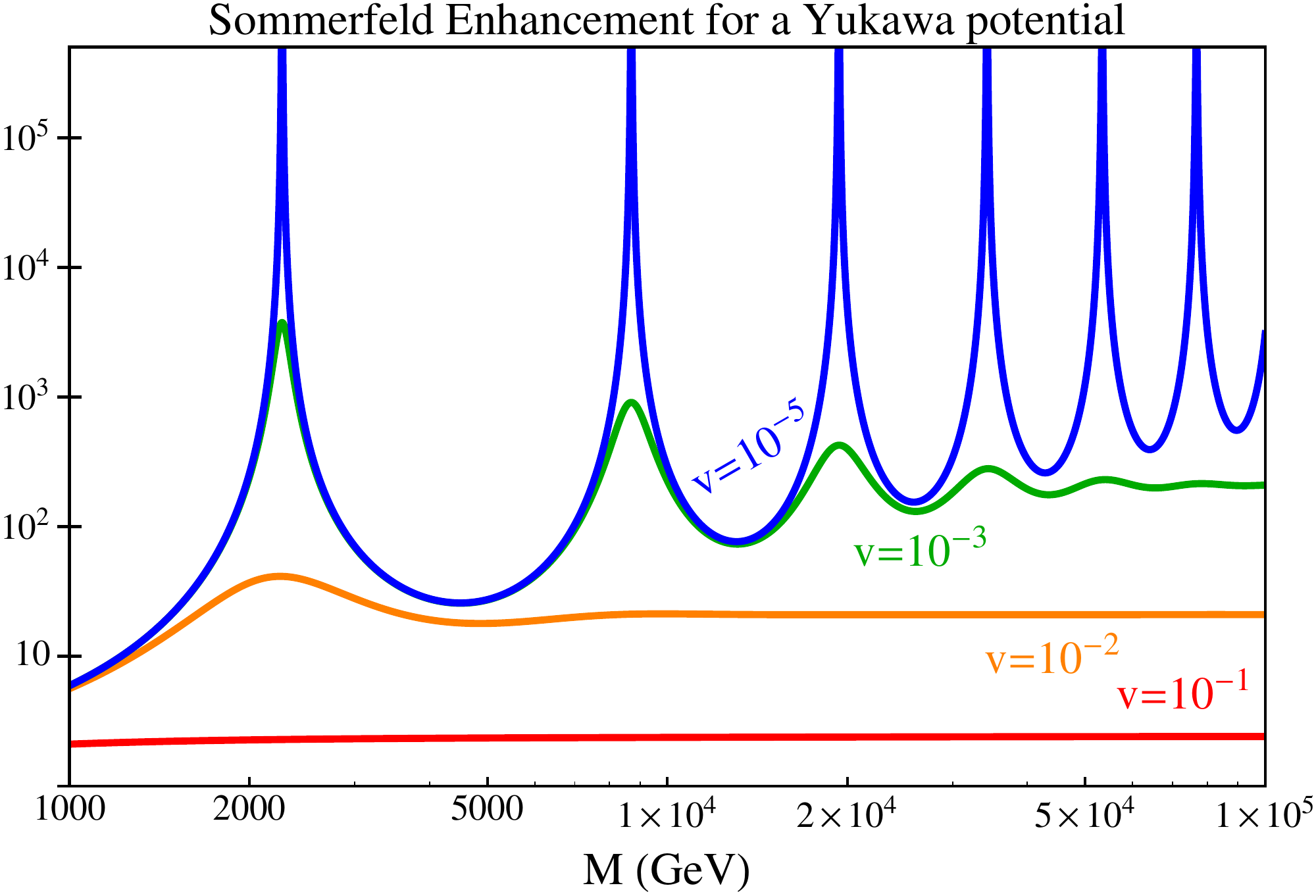}
\caption{Numerical evaluation of the Sommerfeld enhancement factor as a function of the dark matter reduced mass $M$ for a range of relative velocities. The mediator mass is fixed to $90$ GeV and $\alpha = 1/30$. }
\label{fig:yuk}
\end{figure}

For the Coulomb potential $V(r)=-\alpha/r$, one can obtain an analytic expression for the Sommerfeld enhancement \cite{Iengo:2009xf, Iengo:2009ni},
\begin{align}
	S_\ell
	& =
	\frac{e^{\pi \alpha/v} \pi \alpha}{v \sinh\left(\pi\alpha/v\right)\ell!^2}
	\prod_{s=1}^{\ell}
	\left( s^2+\frac{\alpha^2}{v^2} \right)
	\approx 
	\frac{2\pi}{\ell!^2}
	\left( \frac{\alpha}{v} \right)^{2\ell+1}
\end{align}
where the approximation holds for large $\alpha/v$. There exists no simple analytical expression for the enhancement from a Yukawa potential $V(r)=-\alpha e^{-\mu r}/r$, but one can easily evaluate it numerically using the method presented, see Fig.~(\ref{fig:yuk}).  The presence of resonances can be explained by bound states \cite{Lattanzi:2008qa}.  

\section{Box approximation}
\label{app:box}

We have shown that bound state resonances can generate large Sommerfeld enhancements. In this appendix we adapt the procedure used in \cite{Lattanzi:2008qa} to quantitatively understand these resonances.
In \cite{Lattanzi:2008qa}, it was shown that the a reasonable approximation for the Yukawa potential is a flat potential well whose width is determined by the characteristic length scale of the interaction, $r_0=1/m_{\varphi}$,
\begin{align}
V_\text{box}(r) = -U_0 \Theta (r_0 -r).
\label{eq:approximation:box:potential}
\end{align}
The depth of the rectangular well $U_0$ is fixed by requiring that the box approximation matches the Yukawa potential at $r=r_0$,
\begin{align}
V_\text{box}(r) = -\frac{\alpha m}{e} \Theta\left(\frac{1}{m}-r\right).
\end{align}
This approximate is constructed to capture only the qualitative behavior of the full potential and is not a detailed matching to an effective theory.  Observe that this analysis agrees with the fact that the Coulomb limit ($m_{\varphi}\to0$) does not have resonances: this potential has no natural length scale for constructing the rectangular well.

\subsection{Application to $V\sim r^{-3}$}

We adapt this procedure to the singular $1/r^3$ potential,
\begin{align}
V(r) = \frac{-\alpha}{f^2}\frac{1}{r^3}.
\label{eq:approximation:full:NR:pot}
\end{align}
The natural length scale of the problem is the dimensionful scale of the coupling, $r_0 = \sqrt{\alpha}/f$. 
In principle there is also a scale set from the exponential term $e^{-m_\varphi r}$, but for \textsc{uv} models with $m_\varphi \ll f$ this contribution is negligible. This reflects the fact that the resonant behavior of singular potentials in this limit do not depend strongly on the specific value of the mediator mass $m_\varphi$. 
%

This simple box potential approximation provides an estimate for the upper bound of Sommerfeld enhancement coming from resonances in a singular potential.
The solution to the $\ell=0$ Schr\"odinger equation inside the box ($r<r_0$) is
\begin{eqnarray}
\left.\phi(pr<pr_0)\right|_p
    =\frac{ \sin \left(\kappa pr \right)}{\kappa},
\end{eqnarray} 
where $\kappa p = \sqrt{p^2+2U_0M}$.
Outside the box, $r>r_0$, there is effective no potential so that
\begin{eqnarray}
\phi(pr>pr_0)_p=C\sin(pr+\delta).
\end{eqnarray}
$C$ is determined by requiring continuity at $r_0$ so that the enhancement is 
\begin{align}
S &= \left[\cos^2\left(\kappa pr_0\right)+\frac{\sin^2\left(\kappa pr_0\right)}{\kappa^2}\right]^{-1}\nonumber\\
&\approx \left[\cos^2\left(r_0\sqrt{2U_0M}\right)+\frac{p^2}{2MU_0}\sin^2\left(r_0\sqrt{2U_0M}\right)\right]^{-1},
\end{align}
where we use the non-relativistic approximation $p^2\ll U_0M$. Observe that the prefactor of the sine term is small so that $S$ becomes large when the cosine vanishes. In other words, this expression maximized when $r_0\sqrt{2U_0M}=(2n+1)\pi/2$ with
\begin{eqnarray}
S_{\text{max}}\approx \frac{2MU_0}{p^2} = \frac{(2n+1)^2\pi^2}{4r_0^2p^2} .
\end{eqnarray}
This peak is exactly the resonance when the pair of dark matter particles forms a bound state. 
Note that this approximation is independent of the depth of the rectangular well, $U_0$.

%
%


It is straightforward to generalize these expressions for an arbitrary orbital angular momenta, $\ell$, by including the angular barrier to the box potential and applying the appropriate boundary conditions. One obtains
\begin{eqnarray}
S_\ell
=
\left(
\frac{\pi \left[(2\ell')!!\right]^2
\tilde\kappa^{2\ell'}
}{
2^{2\ell'+1}
\Gamma(\ell'+1)^2
}
\right)
\frac{
\left[Y_{\ell'}(pr_0)-\cot(\delta)J_{\ell'}(p r_0)\right]^2
}{
\left[1+\cot^2(\delta)\right]
J^2_{\ell'}(\tilde \kappa p r_0)}
\end{eqnarray}
where $\ell' = \ell + \frac{1}{2}$ and $\tilde\kappa^2 = 2MU_0/p^2$.
The qualitative scaling behavior of the resonance can be seen by setting $\cot(\delta)=0$, and assuming that $pr_0\ll 1$ so that
\begin{eqnarray}
S_\text{max}&\sim& \frac{1}{\left(p^2r_0^2\right)^{2\ell+1}} \sim \frac{1}{v^{4\ell+2}}.
\end{eqnarray}

\subsection{Dimensional analysis}

To estimate the Sommerfeld enhancement off resonance one must estimate $U_0$.  We use the assumption that the \textsc{uv} physics encoded in $U_0$ does not significantly change the \textsc{ir} potential so that the height of the square well $U_0$ is well approximated by the value of the singular potential at the cutoff scale,
\begin{eqnarray}
U_0\sim \frac{f}{\alpha^{1/{2}}} \sim \frac{1}{r_0},
\end{eqnarray}
so that for $\ell=0$, the Sommerfeld enhancement is approximately
\begin{eqnarray}
S\approx\left[\cos^2\left(\sqrt{\frac{2M\alpha^{1/2}}{f}}\right)+
\frac{p^2\alpha^{1/2}}{2Mf}
\sin^2\left(\sqrt{\frac{2M\alpha^{1/2}}{f}}\right)
\right]^{-1}.
\end{eqnarray}
An estimate for the parameters required to hit a resonance without tuning is thus
\begin{eqnarray}
M_\text{res}\sim\frac{1}{r_0}\sim \frac{f}{\alpha^{1/2}},
\end{eqnarray}
which, for most cases, lies at the boundary of the range of the theory's validity.

%
%
%
%
%

\bibliographystyle{utphys} 
\bibliography{DarkMatter}

\providecommand{\href}[2]{#2}\begingroup\raggedright\begin{thebibliography}{10}

\bibitem{Adriani:2008zr}
{\bfseries PAMELA} Collaboration, O.~Adriani {\em et~al.}, ``{An anomalous
  positron abundance in cosmic rays with energies 1.5.100 GeV},''
  \href{http://dx.doi.org/10.1038/nature07942}{{\em Nature} {\bfseries 458}
  (2009) 607--609},
\href{http://arxiv.org/abs/0810.4995}{{\ttfamily arXiv:0810.4995 [astro-ph]}}.

\bibitem{FermiLAT:2011ab}
{\bfseries Fermi LAT Collaboration} Collaboration, M.~Ackermann {\em et~al.},
  ``{Measurement of separate cosmic-ray electron and positron spectra with the
  Fermi Large Area Telescope},''
  \href{http://dx.doi.org/10.1103/PhysRevLett.108.011103}{{\em Phys.Rev.Lett.}
  {\bfseries 108} (2012) 011103},
\href{http://arxiv.org/abs/1109.0521}{{\ttfamily arXiv:1109.0521
  [astro-ph.HE]}}.

\bibitem{Aguilar:2013qda}
{\bfseries AMS Collaboration} Collaboration, M.~Aguilar {\em et~al.}, ``{First
  Result from the Alpha Magnetic Spectrometer on the International Space
  Station: Precision Measurement of the Positron Fraction in Primary Cosmic
  Rays of 0.5–350 GeV},''
\href{http://dx.doi.org/10.1103/PhysRevLett.110.141102}{{\em Phys.Rev.Lett.}
  {\bfseries 110} no.~14, (2013) 141102}.

\bibitem{Weniger:2012tx}
C.~Weniger, ``{A Tentative Gamma-Ray Line from Dark Matter Annihilation at the
  Fermi Large Area Telescope},''
\href{http://arxiv.org/abs/1204.2797}{{\ttfamily arXiv:1204.2797 [hep-ph]}}.

\bibitem{Su:2012ft}
M.~Su and D.~P. Finkbeiner, ``{Strong Evidence for Gamma-ray Line Emission from
  the Inner Galaxy},''
\href{http://arxiv.org/abs/1206.1616}{{\ttfamily arXiv:1206.1616
  [astro-ph.HE]}}.

\bibitem{Rajaraman:2012db}
A.~Rajaraman, T.~M. Tait, and D.~Whiteson, ``{Two Lines or Not Two Lines? That
  is the Question of Gamma Ray Spectra},''
  \href{http://dx.doi.org/10.1088/1475-7516/2012/09/003}{{\em JCAP} {\bfseries
  1209} (2012) 003},
\href{http://arxiv.org/abs/1205.4723}{{\ttfamily arXiv:1205.4723 [hep-ph]}}.

\bibitem{Bloom:2013mwa}
{\bfseries On Behalf of the Fermi-LAT Collaboration} Collaboration, E.~Bloom
  {\em et~al.}, ``{Search of the Earth Limb Fermi Data and Non-Galactic Center
  Region Fermi Data for Signs of Narrow Lines},''
\href{http://arxiv.org/abs/1303.2733}{{\ttfamily arXiv:1303.2733
  [astro-ph.HE]}}.

\bibitem{Ackermann:2012qk}
{\bfseries LAT Collaboration} Collaboration, M.~Ackermann {\em et~al.},
  ``{Fermi LAT Search for Dark Matter in Gamma-ray Lines and the Inclusive
  Photon Spectrum},'' \href{http://dx.doi.org/10.1103/PhysRevD.86.022002}{{\em
  Phys.Rev.} {\bfseries D86} (2012) 022002},
\href{http://arxiv.org/abs/1205.2739}{{\ttfamily arXiv:1205.2739
  [astro-ph.HE]}}.

\bibitem{Iengo:2009ni}
R.~Iengo, ``{Sommerfeld enhancement: General results from field theory
  diagrams},'' \href{http://dx.doi.org/10.1088/1126-6708/2009/05/024}{{\em
  JHEP} {\bfseries 0905} (2009) 024},
\href{http://arxiv.org/abs/0902.0688}{{\ttfamily arXiv:0902.0688 [hep-ph]}}.

\bibitem{Iengo:2009xf}
R.~Iengo, ``{Sommerfeld enhancement for a Yukawa potential},''
\href{http://arxiv.org/abs/0903.0317}{{\ttfamily arXiv:0903.0317 [hep-ph]}}.

\bibitem{Cassel:2009wt}
S.~Cassel, ``{Sommerfeld factor for arbitrary partial wave processes},''
  \href{http://dx.doi.org/10.1088/0954-3899/37/10/105009}{{\em J.Phys.}
  {\bfseries G37} (2010) 105009},
\href{http://arxiv.org/abs/0903.5307}{{\ttfamily arXiv:0903.5307 [hep-ph]}}.

\bibitem{Hisano:2004ds}
J.~Hisano, S.~Matsumoto, M.~M. Nojiri, and O.~Saito, ``{Non-perturbative effect
  on dark matter annihilation and gamma ray signature from galactic center},''
  \href{http://dx.doi.org/10.1103/PhysRevD.71.063528}{{\em Phys.Rev.}
  {\bfseries D71} (2005) 063528},
\href{http://arxiv.org/abs/hep-ph/0412403}{{\ttfamily arXiv:hep-ph/0412403
  [hep-ph]}}.

\bibitem{ArkaniHamed:2008qn}
N.~Arkani-Hamed, D.~P. Finkbeiner, T.~R. Slatyer, and N.~Weiner, ``{A Theory of
  Dark Matter},'' \href{http://dx.doi.org/10.1103/PhysRevD.79.015014}{{\em
  Phys.Rev.} {\bfseries D79} (2009) 015014},
\href{http://arxiv.org/abs/0810.0713}{{\ttfamily arXiv:0810.0713 [hep-ph]}}.

\bibitem{Lattanzi:2008qa}
M.~Lattanzi and J.~I. Silk, ``{Can the WIMP annihilation boost factor be
  boosted by the Sommerfeld enhancement?},''
  \href{http://dx.doi.org/10.1103/PhysRevD.79.083523}{{\em Phys.Rev.}
  {\bfseries D79} (2009) 083523},
\href{http://arxiv.org/abs/0812.0360}{{\ttfamily arXiv:0812.0360 [astro-ph]}}.

\bibitem{Braaten:2013tza}
E.~Braaten and H.~W. Hammer, ``{Universal Two-body Physics in Dark Matter near
  an S-wave Resonance},''
\href{http://arxiv.org/abs/1303.4682}{{\ttfamily arXiv:1303.4682 [hep-ph]}}.

\bibitem{Spergel:1999mh}
D.~N. Spergel and P.~J. Steinhardt, ``{Observational evidence for
  self-interacting cold dark matter},''
  \href{http://dx.doi.org/10.1103/PhysRevLett.84.3760}{{\em Phys. Rev. Lett.}
  {\bfseries 84} (2000) 3760--3763},
\href{http://arxiv.org/abs/astro-ph/9909386}{{\ttfamily
  arXiv:astro-ph/9909386}}.

\bibitem{Dave:2000ar}
R.~Dave, D.~N. Spergel, P.~J. Steinhardt, and B.~D. Wandelt, ``{Halo properties
  in cosmological simulations of selfinteracting cold dark matter},''
  \href{http://dx.doi.org/10.1086/318417}{{\em Astrophys.J.} {\bfseries 547}
  (2001) 574--589},
\href{http://arxiv.org/abs/astro-ph/0006218}{{\ttfamily arXiv:astro-ph/0006218
  [astro-ph]}}.

\bibitem{Rocha:2012jg}
M.~Rocha, A.~H. Peter, J.~S. Bullock, M.~Kaplinghat, S.~Garrison-Kimmel, {\em
  et~al.}, ``{Cosmological Simulations with Self-Interacting Dark Matter I:
  Constant Density Cores and Substructure},''
  \href{http://dx.doi.org/10.1093/mnras/sts514}{{\em Mon.Not.Roy.Astron.Soc.}
  {\bfseries 430} (2013) 81--104},
\href{http://arxiv.org/abs/1208.3025}{{\ttfamily arXiv:1208.3025
  [astro-ph.CO]}}.

\bibitem{Peter:2012jh}
A.~H. Peter, M.~Rocha, J.~S. Bullock, and M.~Kaplinghat, ``{Cosmological
  Simulations with Self-Interacting Dark Matter II: Halo Shapes vs.
  Observations},''
\href{http://arxiv.org/abs/1208.3026}{{\ttfamily arXiv:1208.3026
  [astro-ph.CO]}}.

\bibitem{Tulin:2013teo}
S.~Tulin, H.-B. Yu, and K.~M. Zurek, ``{Beyond Collisionless Dark Matter:
  Particle Physics Dynamics for Dark Matter Halo Structure},''
\href{http://arxiv.org/abs/1302.3898}{{\ttfamily arXiv:1302.3898 [hep-ph]}}.

\bibitem{Navarro:1996bv}
J.~F. Navarro, V.~R. Eke, and C.~S. Frenk, ``{The cores of dwarf galaxy
  halos},''
\href{http://arxiv.org/abs/astro-ph/9610187}{{\ttfamily arXiv:astro-ph/9610187
  [astro-ph]}}.

\bibitem{Chu:2011be}
X.~Chu, T.~Hambye, and M.~H. Tytgat, ``{The Four Basic Ways of Creating Dark
  Matter Through a Portal},''
  \href{http://dx.doi.org/10.1088/1475-7516/2012/05/034}{{\em JCAP} {\bfseries
  1205} (2012) 034},
\href{http://arxiv.org/abs/1112.0493}{{\ttfamily arXiv:1112.0493 [hep-ph]}}.

\bibitem{Buckley:2009in}
M.~R. Buckley and P.~J. Fox, ``{Dark Matter Self-Interactions and Light Force
  Carriers},'' \href{http://dx.doi.org/10.1103/PhysRevD.81.083522}{{\em
  Phys.Rev.} {\bfseries D81} (2010) 083522},
\href{http://arxiv.org/abs/0911.3898}{{\ttfamily arXiv:0911.3898 [hep-ph]}}.

\bibitem{Feng:2009hw}
J.~L. Feng, M.~Kaplinghat, and H.-B. Yu, ``{Halo Shape and Relic Density
  Exclusions of Sommerfeld-Enhanced Dark Matter Explanations of Cosmic Ray
  Excesses},'' \href{http://dx.doi.org/10.1103/PhysRevLett.104.151301}{{\em
  Phys.Rev.Lett.} {\bfseries 104} (2010) 151301},
\href{http://arxiv.org/abs/0911.0422}{{\ttfamily arXiv:0911.0422 [hep-ph]}}.

\bibitem{Georgi:2007ek}
H.~Georgi, ``{Unparticle physics},''
  \href{http://dx.doi.org/10.1103/PhysRevLett.98.221601}{{\em Phys.Rev.Lett.}
  {\bfseries 98} (2007) 221601},
\href{http://arxiv.org/abs/hep-ph/0703260}{{\ttfamily arXiv:hep-ph/0703260
  [hep-ph]}}.

\bibitem{Georgi:2009xq}
H.~Georgi and Y.~Kats, ``{Unparticle self-interactions},''
  \href{http://dx.doi.org/10.1007/JHEP02(2010)065}{{\em JHEP} {\bfseries 1002}
  (2010) 065},
\href{http://arxiv.org/abs/0904.1962}{{\ttfamily arXiv:0904.1962 [hep-ph]}}.

\bibitem{Bellazzini:2011et}
B.~Bellazzini, C.~Csaki, J.~Hubisz, J.~Shao, and P.~Tanedo, ``{Goldstone
  Fermion Dark Matter},''
\href{http://arxiv.org/abs/1106.2162}{{\ttfamily arXiv:1106.2162 [hep-ph]}}.

\bibitem{Chang:2010en}
S.~Chang, N.~Weiner, and I.~Yavin, ``{Magnetic Inelastic Dark Matter},''
  \href{http://dx.doi.org/10.1103/PhysRevD.82.125011}{{\em Phys.Rev.}
  {\bfseries D82} (2010) 125011},
\href{http://arxiv.org/abs/1007.4200}{{\ttfamily arXiv:1007.4200 [hep-ph]}}.

\bibitem{Ferrer:1998rw}
F.~Ferrer and M.~Nowakowski, ``{Higgs and Goldstone bosons mediated long range
  forces},'' \href{http://dx.doi.org/10.1103/PhysRevD.59.075009}{{\em
  Phys.Rev.} {\bfseries D59} (1999) 075009},
\href{http://arxiv.org/abs/hep-ph/9810550}{{\ttfamily arXiv:hep-ph/9810550
  [hep-ph]}}.

\bibitem{Hsu:1992tg}
S.~D. Hsu and P.~Sikivie, ``{Long range forces from two neutrino exchange
  revisited},'' \href{http://dx.doi.org/10.1103/PhysRevD.49.4951}{{\em
  Phys.Rev.} {\bfseries D49} (1994) 4951--4953},
\href{http://arxiv.org/abs/hep-ph/9211301}{{\ttfamily arXiv:hep-ph/9211301
  [hep-ph]}}.

\bibitem{Feinberg:1989ps}
G.~Feinberg, J.~Sucher, and C.~Au, ``{The dispersion theory of dispersion
  forces},''
\href{http://dx.doi.org/10.1016/0370-1573(89)90111-7}{{\em Phys.Rept.}
  {\bfseries 180} (1989) 83}.

\bibitem{Dobrescu:2006au}
B.~A. Dobrescu and I.~Mocioiu, ``{Spin-dependent macroscopic forces from new
  particle exchange},''
  \href{http://dx.doi.org/10.1088/1126-6708/2006/11/005}{{\em JHEP} {\bfseries
  0611} (2006) 005},
\href{http://arxiv.org/abs/hep-ph/0605342}{{\ttfamily arXiv:hep-ph/0605342
  [hep-ph]}}.

\bibitem{Bedaque:2009ri}
P.~F. Bedaque, M.~I. Buchoff, and R.~K. Mishra, ``{Sommerfeld enhancement from
  Goldstone pseudo-scalar exchange},''
  \href{http://dx.doi.org/10.1088/1126-6708/2009/11/046}{{\em JHEP} {\bfseries
  0911} (2009) 046},
\href{http://arxiv.org/abs/0907.0235}{{\ttfamily arXiv:0907.0235 [hep-ph]}}.

\bibitem{Fan:2010gt}
J.~Fan, M.~Reece, and L.-T. Wang, ``{Non-relativistic effective theory of dark
  matter direct detection},''
  \href{http://dx.doi.org/10.1088/1475-7516/2010/11/042}{{\em JCAP} {\bfseries
  1011} (2010) 042},
\href{http://arxiv.org/abs/1008.1591}{{\ttfamily arXiv:1008.1591 [hep-ph]}}.

\bibitem{Fitzpatrick:2012ix}
A.~L. Fitzpatrick, W.~Haxton, E.~Katz, N.~Lubbers, and Y.~Xu, ``{The Effective
  Field Theory of Dark Matter Direct Detection},''
  \href{http://dx.doi.org/10.1088/1475-7516/2013/02/004}{{\em JCAP} {\bfseries
  1302} (2013) 004},
\href{http://arxiv.org/abs/1203.3542}{{\ttfamily arXiv:1203.3542 [hep-ph]}}.

\bibitem{Frank:1971xx}
W.~Frank, D.~Land, and R.~Spector, ``{Singular potentials},''
\href{http://dx.doi.org/10.1103/RevModPhys.43.36}{{\em Rev.Mod.Phys.}
  {\bfseries 43} (1971) 36--98}.

\bibitem{Lepage:1997cs}
G.~Lepage, ``{How to renormalize the Schrodinger equation},''
\href{http://arxiv.org/abs/nucl-th/9706029}{{\ttfamily arXiv:nucl-th/9706029
  [nucl-th]}}.

\bibitem{Beane:2000wh}
S.~Beane, P.~F. Bedaque, L.~Childress, A.~Kryjevski, J.~McGuire, {\em et~al.},
  ``{Singular potentials and limit cycles},''
  \href{http://dx.doi.org/10.1103/PhysRevA.64.042103}{{\em Phys.Rev.}
  {\bfseries A64} (2001) 042103},
\href{http://arxiv.org/abs/quant-ph/0010073}{{\ttfamily arXiv:quant-ph/0010073
  [quant-ph]}}.

\bibitem{Oh:2010ea}
S.-H. Oh, W.~de~Blok, E.~Brinks, F.~Walter, and J.~Kennicutt, Robert~C.,
  ``{Dark and luminous matter in THINGS dwarf galaxies},''
\href{http://arxiv.org/abs/1011.0899}{{\ttfamily arXiv:1011.0899
  [astro-ph.CO]}}.

\bibitem{deNaray:2011hy}
R.~K. de~Naray and K.~Spekkens, ``{Do Baryons Alter the Halos of Low Surface
  Brightness Galaxies?},'' {\em Astrophys.J.} {\bfseries 741} (2011) L29,
\href{http://arxiv.org/abs/1109.1288}{{\ttfamily arXiv:1109.1288
  [astro-ph.CO]}}.

\bibitem{Flores:1994gz}
R.~A. Flores and J.~R. Primack, ``{Observational and theoretical constraints on
  singular dark matter halos},'' {\em Astrophys.J.} {\bfseries 427} (1994)
  L1--4,
\href{http://arxiv.org/abs/astro-ph/9402004}{{\ttfamily arXiv:astro-ph/9402004
  [astro-ph]}}.

\bibitem{Simon:2004sr}
J.~D. Simon, A.~D. Bolatto, A.~Leroy, L.~Blitz, and E.~L. Gates,
  ``{High-resolution measurements of the halos of four dark matter-dominated
  galaxies: Deviations from a universal density profile},''
  \href{http://dx.doi.org/10.1086/427684}{{\em Astrophys.J.} {\bfseries 621}
  (2005) 757--776},
\href{http://arxiv.org/abs/astro-ph/0412035}{{\ttfamily arXiv:astro-ph/0412035
  [astro-ph]}}.

\bibitem{blumenthal1986contraction}
G.~R. Blumenthal, S.~Faber, R.~Flores, and J.~R. Primack, ``Contraction of dark
  matter galactic halos due to baryonic infall,'' {\em The Astrophysical
  Journal} {\bfseries 301} (1986) 27--34.

\bibitem{Gnedin:2004cx}
O.~Y. Gnedin, A.~V. Kravtsov, A.~A. Klypin, and D.~Nagai, ``{Response of dark
  matter halos to condensation of baryons: Cosmological simulations and
  improved adiabatic contraction model},''
  \href{http://dx.doi.org/10.1086/424914}{{\em Astrophys.J.} {\bfseries 616}
  (2004) 16--26},
\href{http://arxiv.org/abs/astro-ph/0406247}{{\ttfamily arXiv:astro-ph/0406247
  [astro-ph]}}.

\bibitem{Tissera:2009cm}
P.~B. Tissera, S.~D. White, S.~Pedrosa, and C.~Scannapieco, ``{Dark matter
  response to galaxy formation},''
\href{http://arxiv.org/abs/0911.2316}{{\ttfamily arXiv:0911.2316
  [astro-ph.CO]}}.

\bibitem{Vogelsberger:2012ku}
M.~Vogelsberger, J.~Zavala, and A.~Loeb, ``{Subhaloes in Self-Interacting
  Galactic Dark Matter Haloes},'' {\em Mon.Not.Roy.Astron.Soc.} {\bfseries 423}
  (2012) 3740,
\href{http://arxiv.org/abs/1201.5892}{{\ttfamily arXiv:1201.5892
  [astro-ph.CO]}}.

\bibitem{Sawala:2010zw}
T.~Sawala, Q.~Guo, C.~Scannapieco, A.~Jenkins, and S.~D. White, ``{What is the
  (Dark) Matter with Dwarf Galaxies?},''
\href{http://arxiv.org/abs/1003.0671}{{\ttfamily arXiv:1003.0671
  [astro-ph.CO]}}.

\bibitem{BoylanKolchin:2011de}
M.~Boylan-Kolchin, J.~S. Bullock, and M.~Kaplinghat, ``{Too big to fail? The
  puzzling darkness of massive Milky Way subhaloes},'' {\em
  Mon.Not.Roy.Astron.Soc.} {\bfseries 415} (2011) L40,
\href{http://arxiv.org/abs/1103.0007}{{\ttfamily arXiv:1103.0007
  [astro-ph.CO]}}.

\bibitem{BoylanKolchin:2011dk}
M.~Boylan-Kolchin, J.~S. Bullock, and M.~Kaplinghat, ``{The Milky Way's bright
  satellites as an apparent failure of LCDM},''
  \href{http://dx.doi.org/10.1111/j.1365-2966.2012.20695.x}{{\em
  Mon.Not.Roy.Astron.Soc.} {\bfseries 422} (2012) 1203--1218},
\href{http://arxiv.org/abs/1111.2048}{{\ttfamily arXiv:1111.2048
  [astro-ph.CO]}}.

\bibitem{Moore:1999nt}
B.~Moore, S.~Ghigna, F.~Governato, G.~Lake, T.~R. Quinn, {\em et~al.}, ``{Dark
  matter substructure within galactic halos},''
  \href{http://dx.doi.org/10.1086/312287}{{\em Astrophys.J.} {\bfseries 524}
  (1999) L19--L22},
\href{http://arxiv.org/abs/astro-ph/9907411}{{\ttfamily arXiv:astro-ph/9907411
  [astro-ph]}}.

\bibitem{Klypin:1999uc}
A.~A. Klypin, A.~V. Kravtsov, O.~Valenzuela, and F.~Prada, ``{Where are the
  missing Galactic satellites?},'' \href{http://dx.doi.org/10.1086/307643}{{\em
  Astrophys.J.} {\bfseries 522} (1999) 82--92},
\href{http://arxiv.org/abs/astro-ph/9901240}{{\ttfamily arXiv:astro-ph/9901240
  [astro-ph]}}.

\bibitem{MiraldaEscude:2000qt}
J.~Miralda-Escude, ``{A test of the collisional dark matter hypothesis from
  cluster lensing},''
\href{http://arxiv.org/abs/astro-ph/0002050}{{\ttfamily arXiv:astro-ph/0002050
  [astro-ph]}}.

\bibitem{Oguri:2010ik}
M.~Oguri, M.~Takada, N.~Okabe, and G.~P. Smith, ``{Direct measurement of dark
  matter halo ellipticity from two-dimensional lensing shear maps of 25 massive
  clusters},'' {\em Mon.Not.Roy.Astron.Soc.} {\bfseries 405} (2010) 2215--2230,
\href{http://arxiv.org/abs/1004.4214}{{\ttfamily arXiv:1004.4214
  [astro-ph.CO]}}.

\bibitem{Finkbeiner:2010sm}
D.~P. Finkbeiner, L.~Goodenough, T.~R. Slatyer, M.~Vogelsberger, and N.~Weiner,
  ``{Consistent Scenarios for Cosmic-Ray Excesses from Sommerfeld-Enhanced Dark
  Matter Annihilation},''
  \href{http://dx.doi.org/10.1088/1475-7516/2011/05/002}{{\em JCAP} {\bfseries
  1105} (2011) 002},
\href{http://arxiv.org/abs/1011.3082}{{\ttfamily arXiv:1011.3082 [hep-ph]}}.

\bibitem{Hannestad:2010zt}
S.~Hannestad and T.~Tram, ``{Sommerfeld Enhancement of DM Annihilation:
  Resonance Structure, Freeze-Out and CMB Spectral Bound},''
  \href{http://dx.doi.org/10.1088/1475-7516/2011/01/016}{{\em JCAP} {\bfseries
  1101} (2011) 016},
\href{http://arxiv.org/abs/1008.1511}{{\ttfamily arXiv:1008.1511
  [astro-ph.CO]}}.

\bibitem{Hisano:2011dc}
J.~Hisano, M.~Kawasaki, K.~Kohri, T.~Moroi, K.~Nakayama, {\em et~al.},
  ``{Cosmological constraints on dark matter models with velocity-dependent
  annihilation cross section},''
  \href{http://dx.doi.org/10.1103/PhysRevD.83.123511}{{\em Phys.Rev.}
  {\bfseries D83} (2011) 123511},
\href{http://arxiv.org/abs/1102.4658}{{\ttfamily arXiv:1102.4658 [hep-ph]}}.

\bibitem{Loeb:2010gj}
A.~Loeb and N.~Weiner, ``{Cores in Dwarf Galaxies from Dark Matter with a
  Yukawa Potential},''
  \href{http://dx.doi.org/10.1103/PhysRevLett.106.171302}{{\em Phys.Rev.Lett.}
  {\bfseries 106} (2011) 171302},
\href{http://arxiv.org/abs/1011.6374}{{\ttfamily arXiv:1011.6374
  [astro-ph.CO]}}.

\bibitem{Bergstrom:2009fa}
L.~Bergstrom, J.~Edsjo, and G.~Zaharijas, ``{Dark matter interpretation of
  recent electron and positron data},''
  \href{http://dx.doi.org/10.1103/PhysRevLett.103.031103}{{\em Phys.Rev.Lett.}
  {\bfseries 103} (2009) 031103},
\href{http://arxiv.org/abs/0905.0333}{{\ttfamily arXiv:0905.0333
  [astro-ph.HE]}}.

\bibitem{Feng:2010zp}
J.~L. Feng, M.~Kaplinghat, and H.-B. Yu, ``{Sommerfeld Enhancements for Thermal
  Relic Dark Matter},''
  \href{http://dx.doi.org/10.1103/PhysRevD.82.083525}{{\em Phys.Rev.}
  {\bfseries D82} (2010) 083525},
\href{http://arxiv.org/abs/1005.4678}{{\ttfamily arXiv:1005.4678 [hep-ph]}}.

\bibitem{Cohen:2012me}
T.~Cohen, M.~Lisanti, T.~R. Slatyer, and J.~G. Wacker, ``{Illuminating the 130
  GeV Gamma Line with Continuum Photons},''
  \href{http://dx.doi.org/10.1007/JHEP10(2012)134}{{\em JHEP} {\bfseries 1210}
  (2012) 134},
\href{http://arxiv.org/abs/1207.0800}{{\ttfamily arXiv:1207.0800 [hep-ph]}}.

\bibitem{Fan:2012gr}
J.~Fan and M.~Reece, ``{A Simple Recipe for the 111 and 128 GeV Lines},''
\href{http://arxiv.org/abs/1209.1097}{{\ttfamily arXiv:1209.1097 [hep-ph]}}.

\bibitem{Jackson:2009kg}
C.~Jackson, G.~Servant, G.~Shaughnessy, T.~M. Tait, and M.~Taoso, ``{Higgs in
  Space!},'' \href{http://dx.doi.org/10.1088/1475-7516/2010/04/004}{{\em JCAP}
  {\bfseries 1004} (2010) 004},
\href{http://arxiv.org/abs/0912.0004}{{\ttfamily arXiv:0912.0004 [hep-ph]}}.

\bibitem{Jackson:2013tca}
C.~Jackson, G.~Servant, G.~Shaughnessy, T.~M.~P. Tait, and M.~Taoso, ``{Gamma
  Rays from Top-Mediated Dark Matter Annihilations},''
\href{http://arxiv.org/abs/1303.4717}{{\ttfamily arXiv:1303.4717 [hep-ph]}}.

\bibitem{Jackson:2013pjq}
C.~Jackson, G.~Servant, G.~Shaughnessy, T.~M.~P. Tait, and M.~Taoso,
  ``{Gamma-ray lines and One-Loop Continuum from s-channel Dark Matter
  Annihilations},''
\href{http://arxiv.org/abs/1302.1802}{{\ttfamily arXiv:1302.1802 [hep-ph]}}.

\bibitem{Goodman:2010qn}
J.~Goodman, M.~Ibe, A.~Rajaraman, W.~Shepherd, T.~M. Tait, {\em et~al.},
  ``{Gamma Ray Line Constraints on Effective Theories of Dark Matter},''
  \href{http://dx.doi.org/10.1016/j.nuclphysb.2010.10.022}{{\em Nucl.Phys.}
  {\bfseries B844} (2011) 55--68},
\href{http://arxiv.org/abs/1009.0008}{{\ttfamily arXiv:1009.0008 [hep-ph]}}.

\end{thebibliography}\endgroup


\end{document}